\documentclass[preprint]{aastex}

\newcommand\arcpt{${{\lower3pt\hbox{$^{\prime\prime}$}}\atop{\raise4pt\hbox{.}}}$}

\slugcomment{to appear in the {\it Astronomical Journal}}

\shorttitle{New Nearby Stars}
\shortauthors{Henry et al.}

\begin{document}

\title{The Solar Neighborhood X: New Nearby Stars in the Southern Sky
and Accurate Photometric Distance Estimates for Red Dwarfs}

\author{Todd J. Henry, John P. Subasavage, Misty A. Brown, Thomas D.
Beaulieu, Wei-Chun Jao}

\affil{Georgia State University, Atlanta, GA 30303--3083}

\author{Nigel C. Hambly}

\affil{Institute for Astronomy, University of Edinburgh}
\affil{Royal Observatory, Blackford Hill, Edinburgh, EH9~3HJ, Scotland, UK}

%\email{thenry@chara.gsu.edu}

%%%%%%%%%%%%%%%%%%%%%%%%%%%%%%%%%%%%%%%%%%%%%%%%%%%%%%%%%%%%%%%%%%%%%%%%%%%%%%
% {Abstract}
%%%%%%%%%%%%%%%%%%%%%%%%%%%%%%%%%%%%%%%%%%%%%%%%%%%%%%%%%%%%%%%%%%%%%%%%%%%%%%

\begin{abstract}

Photometric ($V_J$$R_C$$I_C$) and spectroscopic (6000$-$9500\AA)
observations of high proper motion stars discovered during the first
phase of the SuperCOSMOS RECONS (SCR) search are used to estimate
accurate distances to eight new nearby red dwarfs, including probable
10 pc sample members SCR 1845-6357 (M8.5V at 4.6 pc), the binary SCR
0630-7643AB (M6.0VJ at 7.0 pc), and SCR 1138-7721 (M5.0V at 9.4 pc).
Distance estimates are determined using a suite of new photometric
color-$M_{Ks}$ relations defined using a robust set of nearby stars
with accurate $VRIJHK_s$ photometry and trigonometric parallaxes.

These relations are utilized, with optical and infrared photometry, to
estimate distances on a uniform system (generally good to 15\%) for
two additional samples of red nearby star candidates: several recently
discovered members of the solar neighborhood, and known faint stars
with proper motions in excess of 1.0\arcsec/yr south of DEC = $-$57.5.
Of those without accurate trigonometric parallax measurements, there
are five stars in the first sample and three in the second that are
likely to be within 10 pc.  The two nearest are SO 0253+1652 (M7.0V at
3.7 pc) and DEN 1048-3956 (M8.5V at 4.5 pc).  When combined with SCR
1845-6357, these three stars together represent the largest increase
in the 5 pc sample in several decades.

Red spectra are presented for the red dwarfs and types are given on
the RECONS standard spectral system.  Red spectra are also given for
two new nearby white dwarfs for which we estimate distances from the
photometry of less than 20 pc --- WD 0141-675 (LHS 145, 9.3 pc) and
SCR 2012-5956 (17.4 pc).  WD 0141-675 brings the total number of
systems nearer than 10 pc discussed in this paper to 12.

\end{abstract}


%%%%%%%%%%%%%%%%%%%%%%%%%%%%%%%%%%%%%%%%%%%%%%%%%%%%%%%%%%%%%%%%%%%%%%%%%%%%%%
\section{Introduction}
%%%%%%%%%%%%%%%%%%%%%%%%%%%%%%%%%%%%%%%%%%%%%%%%%%%%%%%%%%%%%%%%%%%%%%%%%%%%%%

Large plate digitization efforts such as the Digitized Sky Survey and
SuperCOSMOS, as well as sky surveys such as 2MASS and SDSS, have led
to a renaissance in the search for faint objects lying undiscovered in
the solar neighborhood.  In particular, the census of the nearest
stars (Henry et al.~1997) is gradually being filled in to the end of
the stellar main sequence.  Many of the new discoveries are the latest
M dwarfs (spectral types M6.0V to M9.5V), some of which are among the
nearest few dozen stellar systems.  Each of the recent discoveries has
been made via high proper motion and/or color surveys, some relying on
the classic work of Luyten (Scholz et al.~2001; Reid \& Cruz 2002),
while others are from entirely original efforts (Delfosse et al.~2001;
Lepine et al.~2002; Teegarden et al.~2003; Cruz et al.~2003; this
work).

Here we report optical photometry and spectroscopy for nine new nearby
star candidates from the first phase of the SuperCOSMOS RECONS (SCR)
search (Hambly et al.~2004) that have proper motions in excess of
1.0\arcsec/yr or have types of M6.0V or later.  One object in
particular, SCR 1845-6357, is remarkable, having $V_J$ = 17.40, $R_C$
= 15.00, $I_C$ = 12.47, and spectral type M8.5V.  We predict that it
lies only 4.6 pc from the Sun, making it the third new stellar system
found within 5 pc in the last few years.  This jump in the 5 pc census
from 44 to 47 systems represents a 7\% increase.  Two of the three new
neighbors are found south of DEC = $-$30, where searches for nearby
stars are less complete than in the north.

We also provide a suite of new, robust, relations using $VRIJHK_s$
photometry to estimate distances for nearby stars.  These relations
hinge on the $M_{Ks}$ magnitudes of single stars within 10 pc with
high quality parallaxes, supplemented with recent results for very red
stars, many of which now have reliable parallaxes placing them within
25 pc (e.g.~Dahn et al.~2002).  When used as an ensemble, these
relations allow photometric distance estimates to be more accurate
than ever before, generally good to 15\%.  We give distance estimates
for the new SCR discoveries, several red dwarfs advertised to be
nearby, and for the sample of known faint stars with $\mu$ $>$
1.0\arcsec/yr south of DEC = $-$57.5.  We also provide a detailed
check of the technique by applying it to the sample of very red dwarfs
with parallaxes.

%%%%%%%%%%%%%%%%%%%%%%%%%%%%%%%%%%%%%%%%%%%%%%%%%%%%%%%%%%%%%%%%%%%%%%%%%%%%%%
\section {Samples}
%%%%%%%%%%%%%%%%%%%%%%%%%%%%%%%%%%%%%%%%%%%%%%%%%%%%%%%%%%%%%%%%%%%%%%%%%%%%%%

Four samples are included in the current effort.  Details are given in
Tables 1 and 2.  The initial phase of the SCR search provides a large
number of new proper motion stars south of DEC = $-$57.5 that have
$R_2$ = 10.0 to 16.5 ($R_2$ represents the photographic $R_{59F}$
band).  Details of the search are given in Hambly et al.~(2004).  In
this paper, we discuss the five SCR stars with $\mu$ $\ge$
1.0\arcsec/yr and four others with $\mu$ = 0.4--1.0\arcsec/yr that
have spectral types of M6.0V or later.  Finder charts are given in
Figure 1.

The second sample includes eight red dwarfs with spectral types M6.0V
to M8.5V that have photometric or spectroscopic distance estimates
bringing them into the sample of the nearest few hundred stellar
systems.  Various distance estimates have been made for these stars
using many techniques, sometimes yielding highly discordant estimates.
Here we provide $VRIJHK_s$ photometry and optical spectroscopy so that
they can be compared on uniform photometric and spectroscopic systems.

The third sample includes previously known stars with $\mu$ $\ge$
1.0\arcsec/yr south of DEC = $-$57.5, the region investigated in the
first phase of the SCR search.  The {\it LHS Catalogue} (Luyten 1979,
hereafter LHS) includes a total of 37 stars meeting these criteria, 22
of which are fainter than $m_R$ = 10 by Luyten's estimate.  Also
meeting these criteria are an additional white dwarf pair found
recently by Scholz et al.~(2001), and a red dwarf found by Pokorny et
al.~(2003), bringing the total sample to 25 objects.  The SCR search
recovered 21 of these --- four were missed due to image blending.

The fourth sample includes 31 M dwarfs with spectral types M6.0V to
M9.5V within 25 pc that are combined with the RECONS sample of stars
within 10 pc to develop high quality color-M$_K$ relations.  We refer
to these stars as the ``supplemental sample''.

%%%%%%%%%%%%%%%%%%%%%%%%%%%%%%%%%%%%%%%%%%%%%%%%%%%%%%%%%%%%%%%%%%%%%%%%%%%%%%
\section {Observations}
%%%%%%%%%%%%%%%%%%%%%%%%%%%%%%%%%%%%%%%%%%%%%%%%%%%%%%%%%%%%%%%%%%%%%%%%%%%%%%
\subsection {Photometry}
%%%%%%%%%%%%%%%%%%%%%%%%%%%%%%%%%%%%%%%%%%%%%%%%%%%%%%%%%%%%%%%%%%%%%%%%%%%%%%

The primary sources for red dwarf photometry in the V$_J$R$_C$I$_C$
system are the comprehensive efforts reported in Bessell (1990),
Bessell (1991), Leggett (1992), Weis (1996), and Dahn et al.~(2002).
Once those sources were exhausted, additional references were used as
listed in Tables 1 and 2, in particular for the reddest M dwarfs.

Optical photometry for the SCR stars and additional objects listed in
Table 1 was obtained in the $V_J$$R_C$$I_C$ bands using the Cerro
Tololo Interamerican Observatory (CTIO) 0.9m telescope during several
observing runs from 2000 to 2004 as part of the NOAO Surveys Program
and the SMARTS (Small and Moderate Aperture Research Telescope System)
Consortium.  The 2048 $\times$ 2046 Tektronix CCD camera was used with
the Tek \#2 $VRI$ filter set.  Standard stars from Graham (1982),
Bessell (1990), and Landolt (1992) were observed through a range of
airmasses each night to place measured fluxes on the $V_J$$R_C$$I_C$
system and to calculate extinction corrections.

Data were reduced using IRAF via typical bias subtraction and dome
flat fielding, utilizing calibration frames taken at the beginning of
each night.  In general, a circular aperture 14\arcsec~in diameter was
used to determine stellar fluxes in order to match the aperture used
by Landolt (1992) for the standard stars.  In cases of crowded fields,
an appropriate aperture 6--12\arcsec~in diameter was used to eliminate
stray light from close sources, and aperture corrections were applied.
Program stars were typically observed on multiple nights, yielding a
measure of the internal, night-to-night, errors of $\pm$0.031, 0.021,
0.025 mag at $V_J$$R_C$$I_C$, respectively, for stars with three or
more nights of data (which are usually the faintest targets).  From
the fits of the standard stars, external errors are estimated to be
$\pm$0.017, 0.015, 0.020 mag $V_J$$R_C$$I_C$, respectively.  From
these two error estimates, we adopt a total error of $\pm$0.03 mag in
each band.  The final magnitudes are given in Table 1.

Infrared photometry in the $JHK_s$ system has been extracted from
2MASS, and is also given in Tables 1 and 2.  The errors are the
$x_msigcom$ errors (where {\it x} is j, h, or k), which give a measure
of the total photometric uncertainty including global and systematic
terms.  The errors are almost always less than 0.05 mag, and are
typically 0.02--0.03 mag.  Notable exceptions are the three white
dwarfs SCR 2012-5956 (errors of 0.05, 0.11, null at $JHK_s$
respectively; at $K_s$ $>$ 15.4, the star is too faint for a reliable
measurement of the magnitude and error), SSPM J2231-7515 (0.04, 0.06,
and 0.12 mag), and SSPM J2231-7514 (0.04, 0.06, and 0.08 mag).

%%%%%%%%%%%%%%%%%%%%%%%%%%%%%%%%%%%%%%%%%%%%%%%%%%%%%%%%%%%%%%%%%%%%%%%%%%%%%%
\subsection {Spectroscopy}
%%%%%%%%%%%%%%%%%%%%%%%%%%%%%%%%%%%%%%%%%%%%%%%%%%%%%%%%%%%%%%%%%%%%%%%%%%%%%%

There are two large, reliable, bodies of spectroscopic work for M
dwarfs that utilize modern CCDs --- that of the RECONS group
(e.g.~Kirkpatrick et al.~1991; Henry et al.~2002) and the Palomar/MSU
(PMSU) survey (Reid et al.~1995; Hawley et al.~1996).  The PMSU types
are based upon comparisons with objects observed by the RECONS group,
but employ a restricted range in wavelength coverage,
$\sim$6300-7200\AA~(for the CTIO setup, which is the one most relevant
for the southern stars discussed here).  Spectral types from other
authors given in Tables 1 and 2 are often the result of comparison to
RECONS spectral types.  In one case (Bessell 1991), spectral types are
on a different system altogether, but a representative spectral type
is better than none.

New spectra were obtained during observing runs in July, October, and
December 2003, and March 2004 at the CTIO 1.5m telescope as part of
the SMARTS Consortium.  The R-C Spectrograph and Loral 1200 X 800 CCD
detector were used with grating \#32, which provided 8.6\AA~resolution
and wavelength coverage from 6000-9500\AA.  In a few cases, spectra
are included from previous CTIO 4.0m runs that utilized the R-C
Spectrograph and Loral 3K $\times$ 1K CCD detector with grating \#181,
which provided 5.7\AA~resolution and wavelength coverage from
5000--10000\AA.  In all cases, data were reduced using IRAF via
typical bias subtraction and dome and/or sky flat fielding, using
calibration frames taken at the beginning of each night.  Fringing at
red wavelengths in the 4.0m data was removed by fitting the fringes
and subtracting them via a tailored IDL program.  Fringing was
effectively removed from the 1.5m data in a more straightforward
manner using a combination of dome and sky flats.

Spectral types were assigned using the ALLSTAR program as described in
Henry et al.~(2002), which currently contains a library of $\sim$500
K5.0V to M9.0V spectra.  RECONS types in Tables 1 and 2 have been
assigned using a new finely-tuned set of M6.0V to M9.0V standards,
illustrated in Figure 2, thereby placing all stars on a uniform
system.  In a few cases, spectral types have shifted from those
previously published by RECONS and others (given in the notes).  These
updates are warranted, because the new spectra reported generally have
higher resolution and broader wavelength coverage than previous
spectra, and were taken on only two telescopes, thereby reducing
instrumental inconsistencies.  Spectra of the eight new SCR red dwarfs
and seven of the eight new very red solar neighbors are shown in
Figures 2--4, ranked from bluest to reddest in each panel.

As predicted by the photometry and the reduced proper motion diagram,
the spectrum of SCR 2012-5956 indicates that it is a white dwarf.  For
comparison, shown in Figure 5 are its spectrum and the similar
spectrum of the white dwarf GJ 440.  An additional white dwarf, LHS
2621, also has a similar spectrum, and to our knowledge this is the
first report that it is a white dwarf.  The lack of any obvious
spectral features indicates that the type could be DC or DQ (because
no strong carbon features typical of DQ white dwarfs are included in
the spectral range shown here; the ``features'' in all three are
telluric).  Also shown is the spectrum of WD 0141-675 (LHS 145), a new
nearby DA white dwarf for which we have determined a parallax within
10 pc (the subject of a future paper), and spectra for two similar DA
white dwarfs (note the defining H$\alpha$ absorption features at
6563\AA~in each of the DA white dwarfs).

%%%%%%%%%%%%%%%%%%%%%%%%%%%%%%%%%%%%%%%%%%%%%%%%%%%%%%%%%%%%%%%%%%%%%%%%%%%%%%
\section {Analysis}
%%%%%%%%%%%%%%%%%%%%%%%%%%%%%%%%%%%%%%%%%%%%%%%%%%%%%%%%%%%%%%%%%%%%%%%%%%%%%%
\subsection {Photometric Distances} 
%%%%%%%%%%%%%%%%%%%%%%%%%%%%%%%%%%%%%%%%%%%%%%%%%%%%%%%%%%%%%%%%%%%%%%%%%%%%%%

To estimate distances to red dwarfs in the three target samples, we
take advantage of recent photometry and trigonometric parallax
determinations, and combine optical $UBVRI$ and near-infrared $JHK_s$
photometry to form an extended baseline over which subtle colors are
evident and additional color ``leverage'' is available.  Using all
eight filter bands provides 28 possible color-$M_{Ks}$ relations.
$M_{Ks}$ has been selected because $K_s$ magnitudes are now available
from 2MASS for nearly every red dwarf considered in nearby star
studies.  These relations can be used in combination to reduce errors
in photometric distance estimates caused by photometric outliers in
the fits and to overcome photometric errors in one or more filters for
the target stars.

The new color-$M_{Ks}$ relations have been developed for red dwarfs
with $M_{Ks}$ $\sim$ 4--11, corresponding to spectral types K0.0V to
M9.5V.  This broad range encompasses eight of the nine SCR stars, all
eight of the recent nearby star candidates, and 19 of the 25 stars in
the known high proper motion sample (the exceptions are generally
white dwarfs).  Two samples of stars --- stars within 10 pc (the
RECONS sample) and the supplemental sample of late-type M dwarfs
within 25 pc --- have been combined to develop reliable relations.
Complete details for the RECONS sample, including all of the
photometric values, will be presented in a future 10 pc summary paper
in this series.  In short, only photometrically single, main sequence
stars within 10 pc are used.  Subdwarfs, close multiple systems and
stars with parallax errors greater than 5 mas (i.e.~5\% errors at
most, at 10 pc) have been removed.  To bolster the red end of the
relations, the RECONS sample has been supplemented with the stars
listed in Table 2.  These stars are nearer than 25 pc, have $M_{Ks}$ =
9 to 11 (spectral types M6.0V to M9.5V), have trigonometric parallaxes
with errors less than 10 mas, are not subdwarfs, and are not known to
be in close multiple systems.  There are few objects of types M7.0V to
M9.5V missing from this list that have published trigonometric
parallaxes larger than 40 mas.

Trigonometric parallaxes have been collected from the two fundamental
parallax references, the Yale Parallax Catalog (van Altena et
al.~1995) and Hipparcos (ESA 1997).  Additional parallaxes have been
determined for many objects with spectral types from M6.0V to M9.5V
during the last decade, primarily through the efforts of Tinney et
al.~(1995), Tinney (1996), and Dahn et al.~(2002).  Hipparcos
parallaxes are available in only a few cases where bright primary
stars could be observed, because all of the stars in Table 2 were too
faint for Hipparcos.  In stellar systems in which the late-type M
dwarf is a companion, parallaxes are included for all components in
the system, thereby taking advantage of many determinations that may
be more accurate than those for the faint component alone.  In one
case, GJ 644C, Soderhjelm (1999) provides an updated Hipparcos
parallax for the primary, GJ 644 ABD, which is actually a close
triple.  When there is more than one parallax determination, the
weighted mean of all available parallaxes for the stellar system has
been adopted, as listed in column 4 of Tables 1 and 2, and column 2 of
Table 4.

Of the 28 possible color-$M_{Ks}$ relations derivable from
$UBVRIJHK_s$, only 12 were deemed useful.  Currently, there is
insufficient reliable U and B photometry for red dwarfs (particularly
past M6.0V), so the 13 relations employing U and B are not used.  In
addition, the three colors derived from $JHK_s$ alone do not have
sufficient baselines to provide reliable distance estimates.  Fits to
each set of color-$M_{Ks}$ data were made for second through eighth
orders.  Overall, fifth-order fits proved reliable for all 12 colors
used, and higher orders did not improve the fits in any meaningful
way.

An exemplary fit, for $M_{Ks}$ vs.~$(V-K_s)$, is shown in Figure 6.
For each of the 12 relations adopted, the applicable color range, the
numbers of stars used from the RECONS and very red samples, the fit
coefficients, and the RMS values of the fits are given in Table 3.
Equations for the relations have the following format:

$M_{Ks}$ $=$ $+$ 0.00959 $(V-K_s)$$^4$ $-$ 0.23953 $(V-K_s)$$^3$ $+$ 2.05071 $(V-K_s)$$^2$ 

\hskip40pt   $-$ 5.98231$(V-K_s)$ $+$ 9.77683

\noindent Of course, trigonometric parallaxes for additional objects
(including many of those investigated here) and photometry would allow
the improvement of the photometric relation matrix, in particular for
the stars with types later than M7.0V.  We note that Dahn et
al.~(2002), who include accurate data for many stars of this type,
provide an $M_J$ relation for 2.8 $<$ $I-J$ $<$ 4.2 (spectral types
M6.5V to L8.0V).

The photometric distance estimates for stars in the four samples are
given in Table 4, where the number of color-$M_{Ks}$ relations used to
generate each mean distance is listed.  To determine the reliability
of these distance estimates, we have run the complete sample of 140
stars used to generate the fits back through the relations (not all
stars have all colors, so 140 exceeds the total number of stars used
in each fit, as listed in Table 3).  The resulting average error of
the suite of relations technique is 15.3\%.  The final errors listed
for the distance estimates throughout this paper include this 15.3\%
error (the ``external'' error from the fits) and the standard
deviation of the up to 12 different distance estimates for an
individual star (the ``internal'' error for each star).

For the supplemental sample, which represents the reddest dwarfs
focused on in this paper, the differences between the photometric and
trigonometric distances shown in Table 4 are remarkably good, with a
mean difference of only 9.3\%.  This is rather better than the 15.3\%
value obtained when considering the entire range of colors covered by
the suite of relations.  There are at least two possible causes ---
either the very red dwarfs are better constrained in M$_{Ks}$ than
their earlier type counterparts, or the lack of data for the reddest
dwarfs currently provides a poorer measure of the spread in M$_{Ks}$
values.  Additional data on the reddest dwarfs (trigonometric
parallaxes in particular) should determine which is the true cause.

Illustrated in Figure 7 is a comparison between the two distance
determinations for the 31 stars in the supplemental sample.  Note that
nearly every point is within 1$\sigma$ of the equal distance line,
indicating the strength of the photometric distance technique.  Either
the photometry or trigonometric parallax (or both) is suspect for the
single clear outlier, ESO 207-61, which has an offset of 37.1\%
between the estimated and true distances.  The remaining 30 stars
(97\%) have distances estimated to better than 30.6\% (two times the
adopted error for the technique for the full ranges of the relations),
and 25 stars (81\%) have distance estimates better than 15.3\%,
thereby indicating that the mean difference is a conservative
representation of a 1$\sigma$ ``error'' for these stars.  As
mentioned, we have vetted the sample for known close multiple systems,
but additional companions may yet be found to some of the stars
included.

For the white dwarfs, we have utilized equation (7) of Salim et
al.~(2004), which relates $M_V$ and $(V-I)$ for white dwarfs from
Bergeron et al.~(2001).  A trial run of 11 white dwarfs in the RECONS
sample with reliable parallaxes and $V_J$$R_C$$I_C$ photometry through
that single relation yields an average difference between the
photometric and trigonometric distances of 13.2\%.  However, there is
significantly more scatter than for results from the red dwarfs'
photometric relation matrix --- three of the 11 white dwarf distance
estimates are discordant by more than 20\%.  Because we do not have
multiple relations to use to determine an individual error for each
white dwarf distance, we conservatively adopt a generic 20\% error for
each estimate (8 of 11, or 73\% of the white dwarfs are within 20\%,
corresponding roughly to a 1$\sigma$ ``error'' for roughly two-thirds
of the stars).  We plan to create a white dwarf photometric relation
matrix and report the results in a future paper in this series.

%%%%%%%%%%%%%%%%%%%%%%%%%%%%%%%%%%%%%%%%%%%%%%%%%%%%%%%%%%%%%%%%%%%%%%%%%%%%%%
\section {Discussion}
%%%%%%%%%%%%%%%%%%%%%%%%%%%%%%%%%%%%%%%%%%%%%%%%%%%%%%%%%%%%%%%%%%%%%%%%%%%%%%

Typically, distance estimates are made for the most compelling new
nearby star candidates, but comparing one discovery to another is
difficult because different methods are used in each publication.  The
methods used to estimate distances include nearly any permutation
involving optical photometry, infrared photometry, optical
spectroscopy, and trigonometric parallaxes.  Oftentimes, only a few
reference stars are used to establish the distance to a newly
discovered star.  In addition, as is evident from the need for 25
different references in Table 2, there is a general lack of
homogeneity in basic information for the reddest stars.

Here we remedy some of these problems by estimating distances to stars
in all three target samples using the suite of relations in a uniform
way that allows accurate distance estimates, and perhaps more
important, allows the target stars to be compared directly.  This
effort can overcome some of the differences in photometry and
spectroscopy inherent to different observers and their techniques.  We
highlight results for noteworthy stars here.

%%%%%%%%%%%%%%%%%%%%%%%%%%%%%%%%%%%%%%%%%%%%%%%%%%%%%%%%%%%%%%%%%%%%%%%%%%%%%%
\subsection {Comments on the SCR Discoveries}
%%%%%%%%%%%%%%%%%%%%%%%%%%%%%%%%%%%%%%%%%%%%%%%%%%%%%%%%%%%%%%%%%%%%%%%%%%%%%%

SCR 0342-6407 ($\mu$ = 1.071\arcsec/yr @ 141.4$^\circ$) is the most
distant of the five SCR discoveries with $\mu$ $>$ 1.0\arcsec/yr at
38.1 $\pm$ 7.8 pc, and has the earliest spectral type, M4.5V.

SCR 0630-7643AB ($\mu$ = 0.483\arcsec/yr @ 356.8$^\circ$) is a close
binary.  Images indicate two sources with a constant separation of
1.0\arcsec~over five months and brightness ratio of 0.8 at $I$.  The
combined photometry yields a distance estimate of 5.2 $\pm$ 0.9 pc.
Assuming a brightness difference of 0.25 mag at all wavelengths (no
color information is currently available), the distance estimate is
7.0 $\pm$ 1.2 pc.  By either estimate, the system is almost certainly
a new member of the RECONS sample and a promising target for future
mass determinations.

SCR 1138-7721 ($\mu$ = 2.141\arcsec/yr @ 286.8$^\circ$) is a
possible new member of the RECONS sample, falling just within the 10
pc horizon at 9.4 $\pm$ 1.7 pc.

SCR 1845-6357 ($\mu$ = 2.558\arcsec/yr @ 074.8$^\circ$) is the third
recent discovery of a late M dwarf that is probably within 5 pc, with
a distance estimate of 4.6 $\pm$ 0.8 pc.  At 2.6\arcsec/yr, it has the
highest proper motion of the 120 new SCR stars discovered south of DEC
= $-$57.5 that have $\mu$ $\ge$ 0.4\arcsec/yr.

SCR 1848-6855 ($\mu$ = 1.287\arcsec/yr @ 194.4$^\circ$) appears to
be a normal M5.0V star with no obvious subdwarf features.  Given its
estimated distance of 37.0 $\pm$ 9.4 pc and high proper motion, it has
a relatively high tangential velocity of 225 km/sec.

SCR 2012-5956 ($\mu$ = 1.440\arcsec/yr @ 165.6$^\circ$) is a new
nearby white dwarf, with an estimated distance of 17.4 pc, using
equation (7) of Salim et al.~(2004).

%%%%%%%%%%%%%%%%%%%%%%%%%%%%%%%%%%%%%%%%%%%%%%%%%%%%%%%%%%%%%%%%%%%%%%%%%%%%%%
\subsection {Comments on the Recently Discovered Nearby Late M Dwarfs}
%%%%%%%%%%%%%%%%%%%%%%%%%%%%%%%%%%%%%%%%%%%%%%%%%%%%%%%%%%%%%%%%%%%%%%%%%%%%%%

SO 0253+1652 has been claimed to be the third nearest star, at a
distance of only 2.4 $\pm$ 0.5 pc based on a crude trigonometric
parallax and 3.6 $\pm$ 0.4 pc photometrically using $V$ = 15.40, $R$
= 13.26, $I$ = 10.66 (Teegarden et al.~2003).  Our photometry from
three nights gives $(V-I)$ = 4.24 rather than their $(V-I)$ = 4.74,
and the relations place it further away, at 3.7 $\pm$ 0.6 pc, than
their trigonometric distance.  This would rank it as the 22nd nearest
system, rather than as the third nearest system indicated by the
poorly determined trigonometric value.

LP 775-31 and LP 655-48 are M7.0V ``twins'' reported by McCaughrean et
al.~(2002) to be 6.2--6.5 pc and 7.9--8.2 pc from the Sun,
respectively, using a combination of spectroscopy and infrared
photometry, although they reported types of M8.0V and M7.5V.  Reid \&
Cruz (2002) estimated distances of 7.4 $\pm$ 1.5 pc and 7.7 $\pm$ 1.5
pc.  Cruz \& Reid (2002) gave types of M6.0V for both and estimated
distances of 11.3 $\pm$ 1.3 pc and 15.3 $\pm$ 2.6 pc.  We confirm that
they are probable new members of the RECONS sample, and provide
distance estimates of 7.3 $\pm$ 1.2 pc and 8.2 $\pm$ 1.4 pc,
respectively.

LHS 2021 is an M7.5V star that had, to our knowledge, no distance
estimate until that given here.  At 13.8 $\pm$ 2.3 pc, it is just
beyond the RECONS sample horizon.

LHS 2090 was reported in Scholz et al.~(2001) to have spectral type
M6.5V and was estimated to be 6.0 $\pm$ 1.1 pc distant based on its
2MASS $JHK_s$ photometry and that of a similar star, GJ 1111.  Reid \&
Cruz (2002) report a distance estimate of 5.2 $\pm$ 1.0 pc, based upon
an ($M_{Ks}$, $J-K_s$) relation.  We derive a distance of 5.7 $\pm$
0.9 pc.

DEN 1048-3956 was reported in Delfosse et al.~(2001) to have spectral
type M9.0V and magnitudes $B$ = 19.0 and $R$ = 15.7 from Schmidt
plates, and $I$ = 12.67, $J$ = 9.59, $K$ = 8.58 from the DENIS survey.
Gizis (2002) determined a spectral type of M8.0V.  Delfosse et al.'s
comparison to four M dwarfs with type M9.0V yielded a distance
estimate of 4.1 $\pm$ 0.6 pc.  Preliminary trigonometric parallaxes
have been reported by Deacon \& Hambly (2001, 5.2 $\pm$ 1.0 pc) and
Neuhauser et al.~(2002, 4.6 $\pm$ 0.3 pc), although the latter uses a
faint set of reference stars, assumes a proper motion rather than
solving for it, and utilizes only three frames.  Our photometric
distance estimate is 4.5 $\pm$ 0.7 pc.

LHS 325a was reported as LHS 325 in Bessell (1991), but it must be in
fact LHS 325a, an insertion in the LHS catalog between LHS 325 and 326
because of its RA.  We have found it to be rather brighter in $VRI$
than Bessell (1991), who gives $V$ = 18.67, $R$ = 16.60, $I$ = 14.36.

LSR 1826+3014 is an M8.5V object found by Lepine et al.~(2002) and
noted to be the faintest ($V$ = 19.36) red dwarf discovered to have a
proper motion larger than 2\arcsec/yr.  Their distance estimate of
13.9 $\pm$ 3.5 pc is based on three photometric/spectroscopic
estimates.  Given the northern declination, we have not observed this
object from CTIO.  Nonetheless, our distance estimate from published
data and 10 relations is 14.5 $\pm$ 2.5 pc, which is certainly
consistent with theirs.

%%%%%%%%%%%%%%%%%%%%%%%%%%%%%%%%%%%%%%%%%%%%%%%%%%%%%%%%%%%%%%%%%%%%%%%%%%%%%%
\subsection {Comments on Additional Objects}
%%%%%%%%%%%%%%%%%%%%%%%%%%%%%%%%%%%%%%%%%%%%%%%%%%%%%%%%%%%%%%%%%%%%%%%%%%%%%%

Among the known high proper motion stars, the true distances to GJ 85
(LHS 150), GJ 181.1 (LHS 199), GJ 1077 (LHS 205), LHS 288, GJ 467AB
(LHS 328 and 329), and GJ 808 (LHS 499) are poorly known, given that
their parallaxes have errors larger than 10 mas.  Not surprisingly,
all but GJ 1077 have photometric distances differing from the
trigonometric distances by more than 20\%.

WD 0141-675 (LHS 145) is a nearby white dwarf with no trigonometric
parallax.  We estimate a distance of 9.3 $\pm$ 1.9 pc using equation
(7) of Salim et al.~(2004).  This is a probable new member of the
RECONS sample and among the 20 nearest known white dwarfs.

GJ 1123 (LHS 263) has a photometric distance estimate of 7.5 $\pm$ 1.2
pc, which matches the spectroscopic estimate of 7.6 pc by Henry et
al.~(2002).  It is a likely new member of the RECONS 10 pc sample.

GJ 1128 (LHS 271) has a photometric distance estimate of 6.4 $\pm$ 1.0
pc, which matches the spectroscopic estimate of 6.6 pc by Henry et
al.~(2002).  It is a likely new member of the RECONS 10 pc sample.

GJ 1277 (LHS 532) is another probable new member of the RECONS sample,
at a distance of 8.9 $\pm$ 1.4 pc.

LP 944-20 is an important nearby red dwarf (or brown dwarf) at the end
of the M spectral sequence.  The first parallax was reported by Tinney
(1996), placing it at a distance of slightly less than 5 pc, and
making it the most recent addition to the 5 pc sample other than GJ
1061 (Henry et al.~1997).  The 22\% difference in the photometric and
trigonometric distances occurs because there are only three colors
available, and all lie near the very limit of each relation.

ESO 207-61 has the poorest match between the photometric and
trigonometric parallaxes, which differ by 37\%.  Ruiz et al.~(1991)
indicate that the $VRI$ photometry is on the Kron-Cousins system, but
the observations were made at the CTIO 0.9m (where our observations
are made) and likely used the same filter set, as well as similar
standards.  Nonetheless, the photometry may not be on the Cousins
system as we have assumed (and reported in Table 2), thereby causing
the discordant distances.

LHS 523 is the final star of the three in the red dwarf supplemental
sample having a mismatch between photometric and trigonometric
distances greater than 20\%, in this case 27\%.  The 12 distance
estimates are the least consistent of all of the stars in the
supplemental sample.  In particular, distances from colors including
the $I$ band range from 9.6 to 20.9 pc, indicating a possible problem
with the $I$ photometry.

%%%%%%%%%%%%%%%%%%%%%%%%%%%%%%%%%%%%%%%%%%%%%%%%%%%%%%%%%%%%%%%%%%%%%%%%%%%%%%
\section {Conclusions}
%%%%%%%%%%%%%%%%%%%%%%%%%%%%%%%%%%%%%%%%%%%%%%%%%%%%%%%%%%%%%%%%%%%%%%%%%%%%%%

The 12 new color-$M_{Ks}$ relations given here can be used to estimate
distances accurately to stars falling in the color ranges given
(generally, K0.0V to M9.5V in spectral type) in any combination.  As
an ensemble, they provide a powerful means to estimate distances to
nearby star candidates on a uniform system.

We predict that the three most compelling targets, SO 0253+1652, DEN
1048-3956, and SCR 1845-6357 all lie within 5 pc.  These three late M
dwarfs comprise the largest surge in the nearby star population in
several decades.

With $\mu$ = 2.6\arcsec/yr, SCR 1845-6357 ranks as the 41st fastest
proper motion system known, so it is not surprising that it is
probably nearer than 5 pc.  Of the 22 other red dwarfs discussed here
that have no accurate trigonometric parallaxes, we find that 10 are
likely to be closer than 10 pc.  In addition, the white dwarf, WD
0141-675, is also probably nearer than 10 pc, bringing the total
number of new 10 pc systems discussed here to 12.  This sample is
currently being observed in the RECONS Cerro Tololo Interamerican
Observatory Parallax Investigation (CTIOPI) carried out at the CTIO
0.9m, which should yield parallaxes for these high priority targets in
the near future.

%%%%%%%%%%%%%%%%%%%%%%%%%%%%%%%%%%%%%%%%%%%%%%%%%%%%%%%%%%%%%%%%%%%%%%%%%%%%%%
\section {Acknowledgments}
%%%%%%%%%%%%%%%%%%%%%%%%%%%%%%%%%%%%%%%%%%%%%%%%%%%%%%%%%%%%%%%%%%%%%%%%%%%%%%

We wish to thank Charlie Finch, Hektor Monteiro, and Jen Winters for
their supporting work in this effort.  The photometric and
spectroscopic observations reported here were carried out under the
auspices of the NOAO Surveys Program and via the SMARTS (Small and
Moderate Aperture Research Telescope System) Consortium, which
operates several small telescopes at CTIO, including the 0.9m and
1.5m.  TJH's Space Interferometry Mission grant supported much of the
work carried out here.  This effort has utilized the SuperCOSMOS, Two
Micron All Sky Survey, and SIMBAD databases.

%%%%%%%%%%%%%%%%%%%%%%%%%%%%%%%%%%%% REFS %%%%%%%%%%%%%%%%%%%%%%%%%%%%%%%%%%%%

\clearpage

%%%%%%%%%%%%%%%%%%%%%%%%%%% FIGURE CAPTIONS %%%%%%%%%%%%%%%%%%%%%%%%%%%%%%%%%%

\figcaption[ToddHenry.f1.eps]{Finder charts for the nine SCR stars from
the SuperCOSMOS red UK Schmidt images (photographic passband
$R_{59F}$).  Each chart is 5\arcmin~on a side, with north up and east
to the left, and the epoch of the image is given.
\label{fig1}}

\figcaption[ToddHenry.f2.eps]{Spectra of standard late-type red dwarfs,
defining each 0.5 subtype from M6.0V to M9.0V.  Important spectral
features are labeled at the top.  The absorption complex at 9300
\AA~and redward is due in part to H$_2$O in the Earth's
atmosphere. \label{fig2}}

\figcaption[ToddHenry.f3.eps]{Spectra of new SCR discoveries and recently
found nearby late M dwarfs with spectral types of M4.5V to M6.0V.
\label{fig3}}

\figcaption[ToddHenry.f4.eps]{Spectra of new SCR discoveries and recently
found nearby late M dwarfs with spectral types of M6.5V to M8.5V.
\label{fig4}}

\figcaption[ToddHenry.f5.eps]{Spectra of the nearby white dwarfs LHS
145 (type DA, note the H$\alpha$ feature at 6563\AA) and SCR 2012-5956
(type DC or DQ).  Also shown for comparison are spectra for white
dwarfs of similar types.  The DC/DQ types are virtually featureless in
this wavelength region --- the ``features'' seen are all telluric.
\label{fig5}}

\figcaption[ToddHenry.f6.eps]{Example color-$M_{Ks}$ fit of stars in
the RECONS and supplemental samples, illustrating the $(V-K_s)$
relation.  Solid points represent RECONS stars within 10 pc.  Open
points represent stars in the supplemental sample of objects with
spectral types of M6.0V and later.  Vertical lines indicate the valid
limits of the relation at $(V-K_s)$ = 2.24 to 9.27. \label{fig6}}

\figcaption[ToddHenry.f7.eps]{Comparison of true trigonometric and
derived photometric distances for the 31 late M dwarfs with accurate
trigonometric parallaxes listed in Tables 2 and 4.  The three labeled
stars are discussed in the text.  \label{fig7}}

%%%%%%%%%%%%%%%%%%%%%%%%%%%%%%%%%%% FIGURES %%%%%%%%%%%%%%%%%%%%%%%%%%%%%%%%%%
														  
\begin{figure}
\plotone{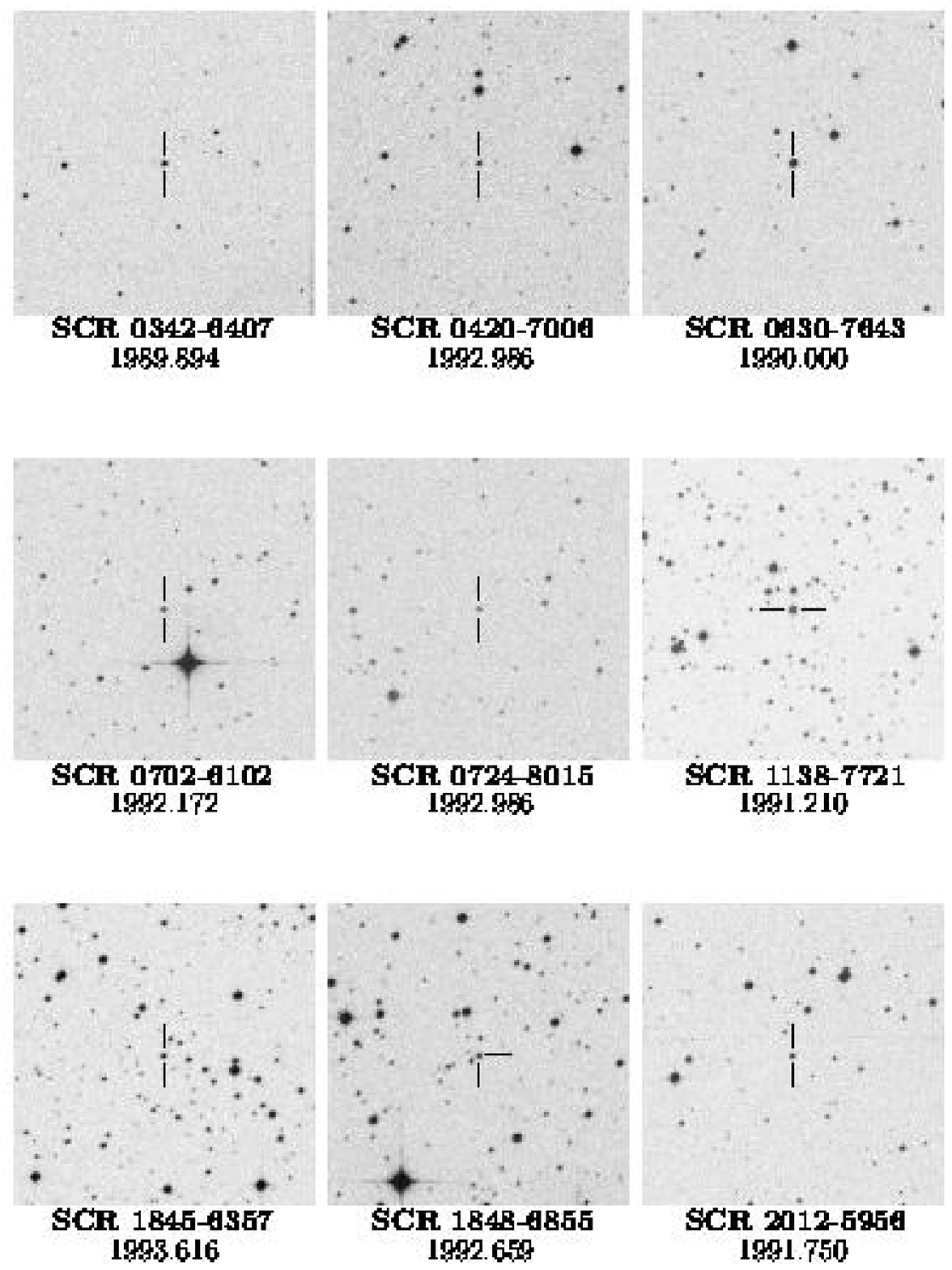}
\label{fig.spec.standards}
\end{figure}

\begin{figure}
\plotone{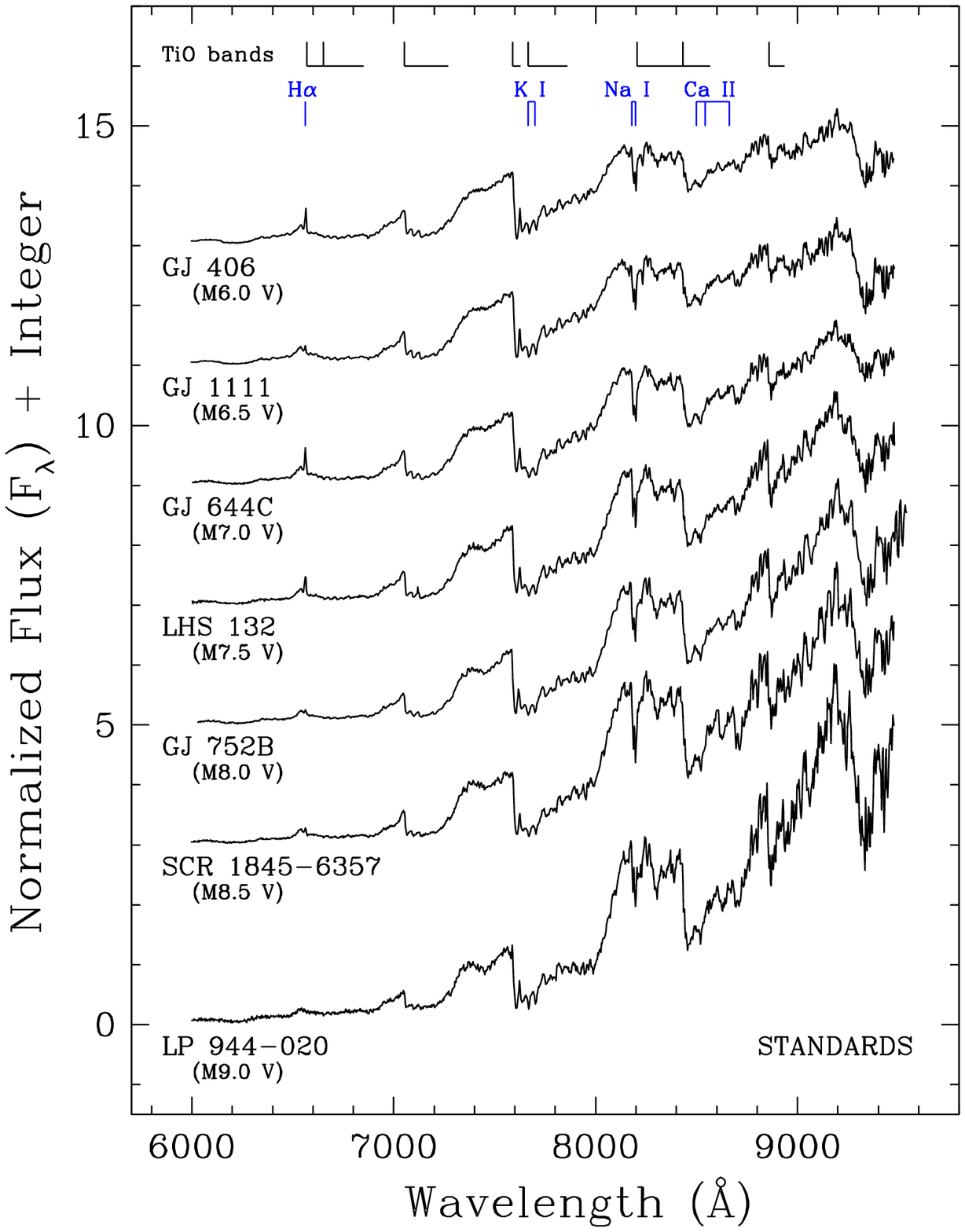}
\label{fig.spec.standards}
\end{figure}

\begin{figure}
\plotone{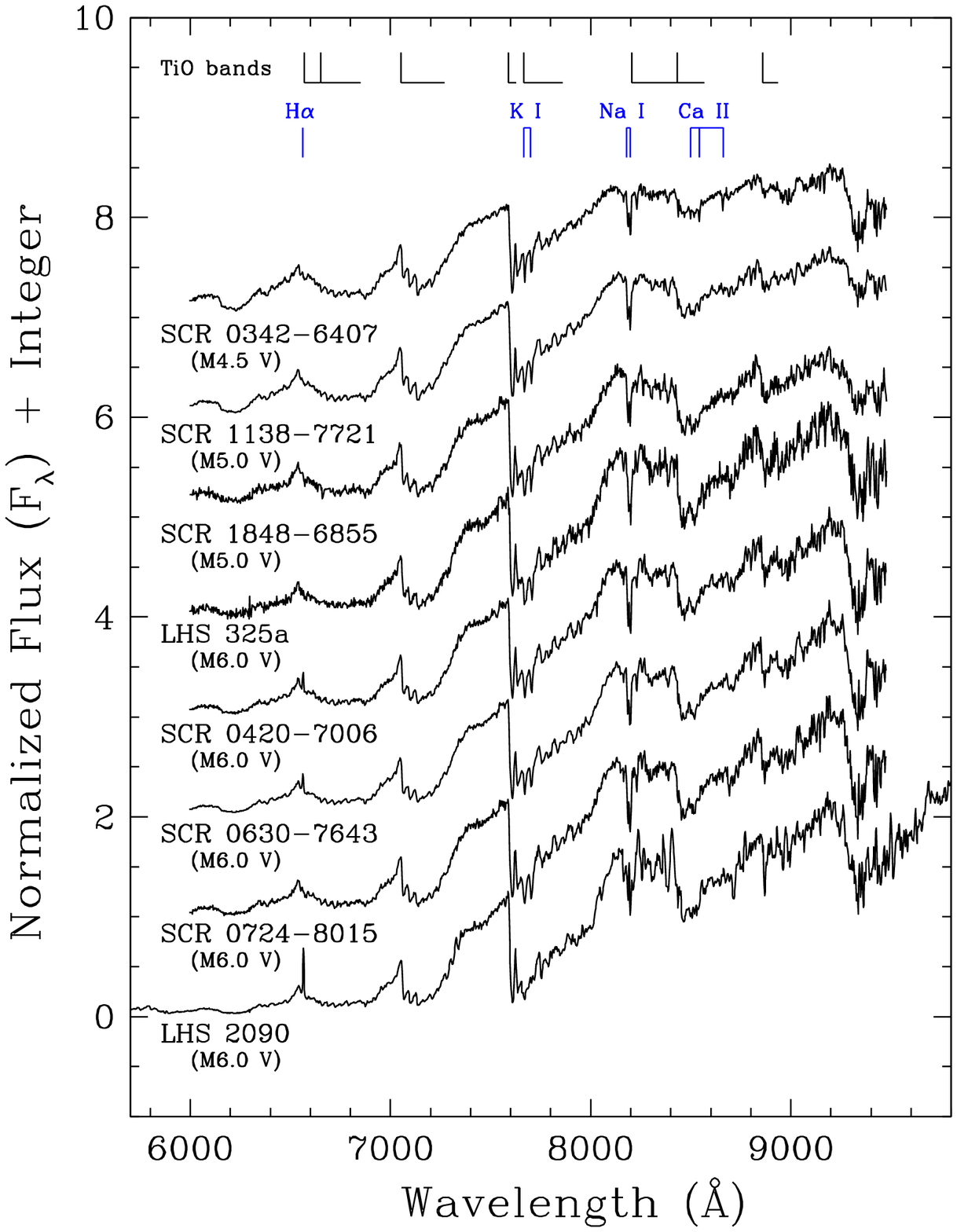}
\label{fig.spec.earlysci}
\end{figure}

\begin{figure}
\plotone{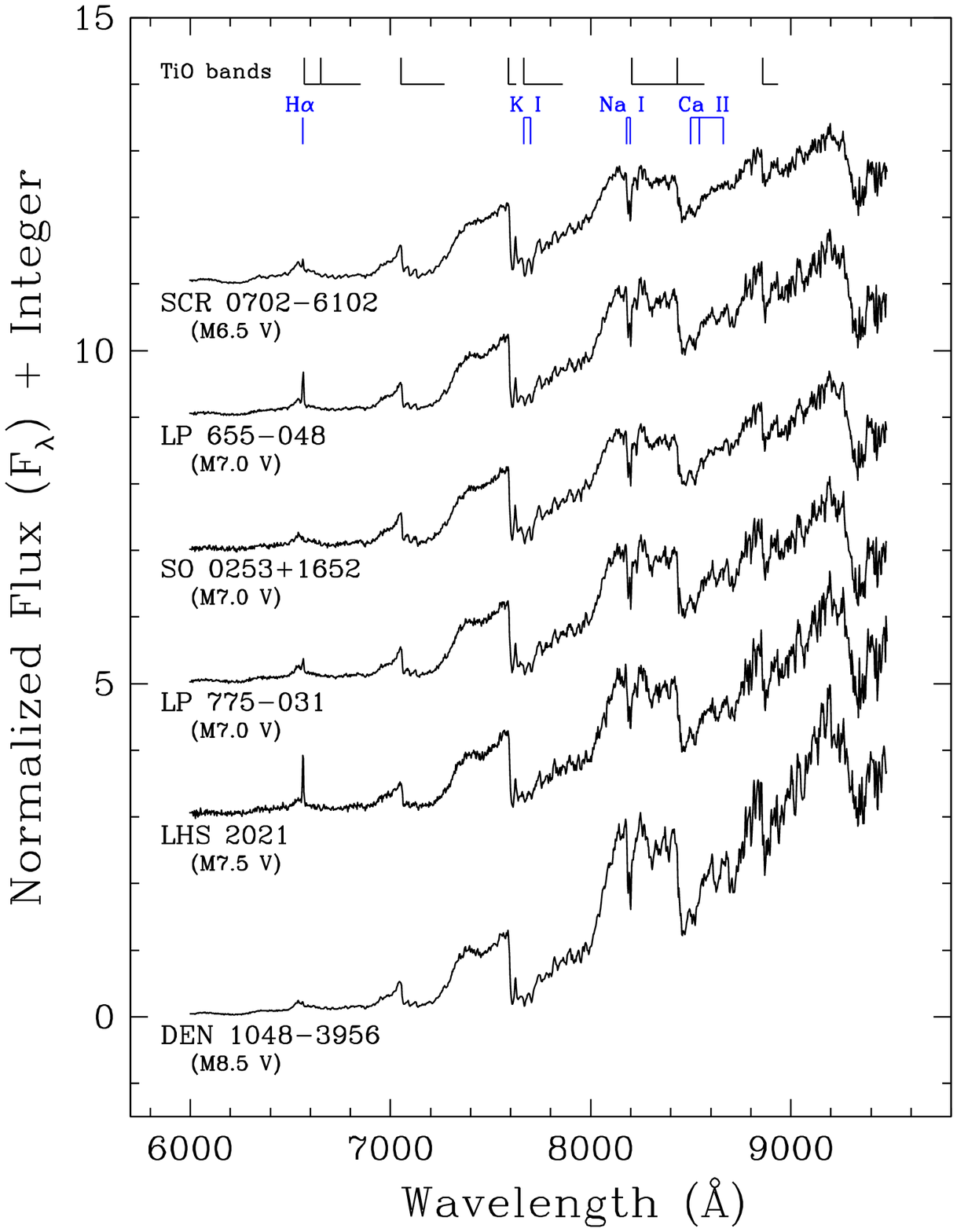}
\label{fig.spec.latesci}
\end{figure}

\begin{figure}
\plotone{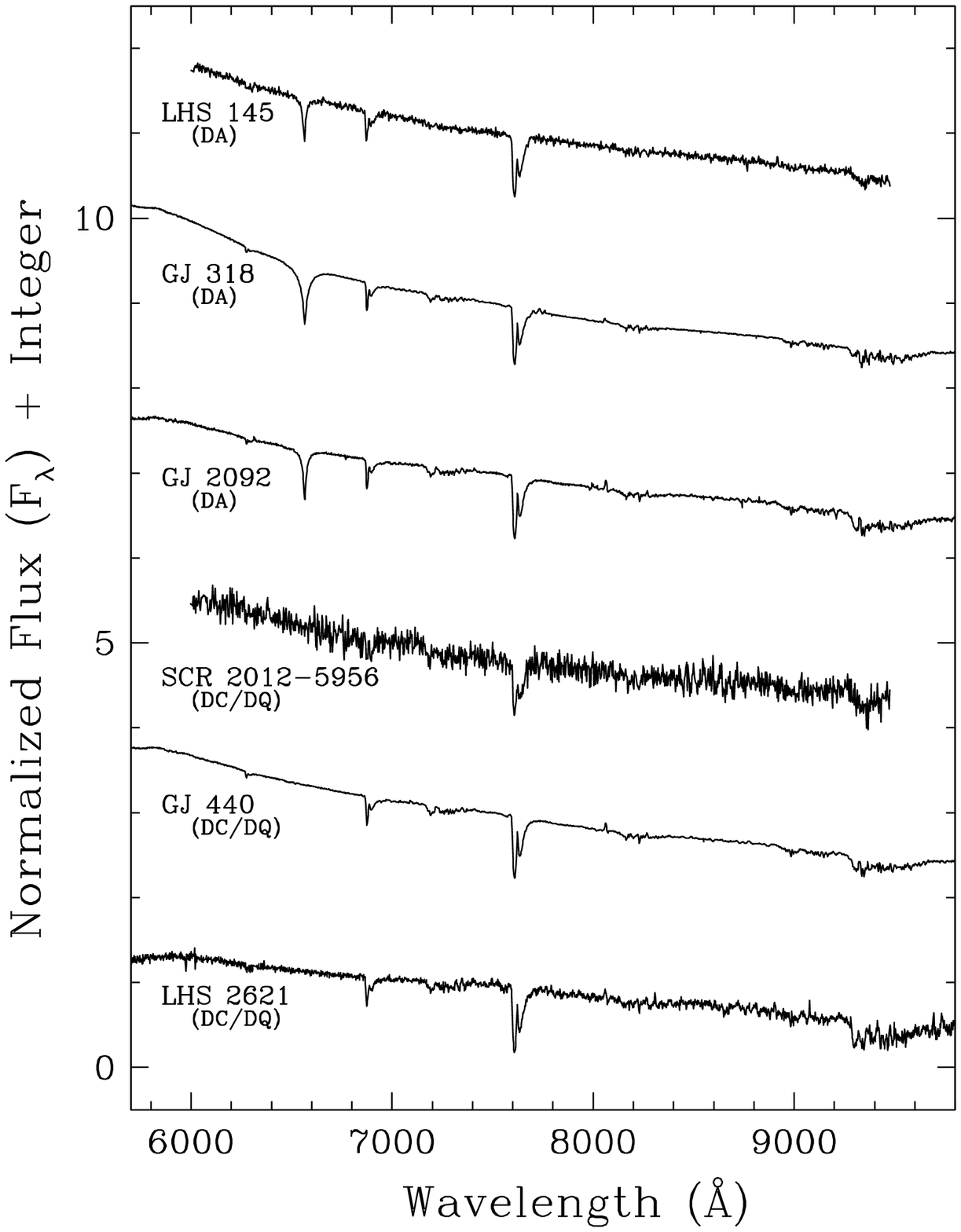}
\label{fig.spec.wd}
\end{figure}

\begin{figure}
\plotone{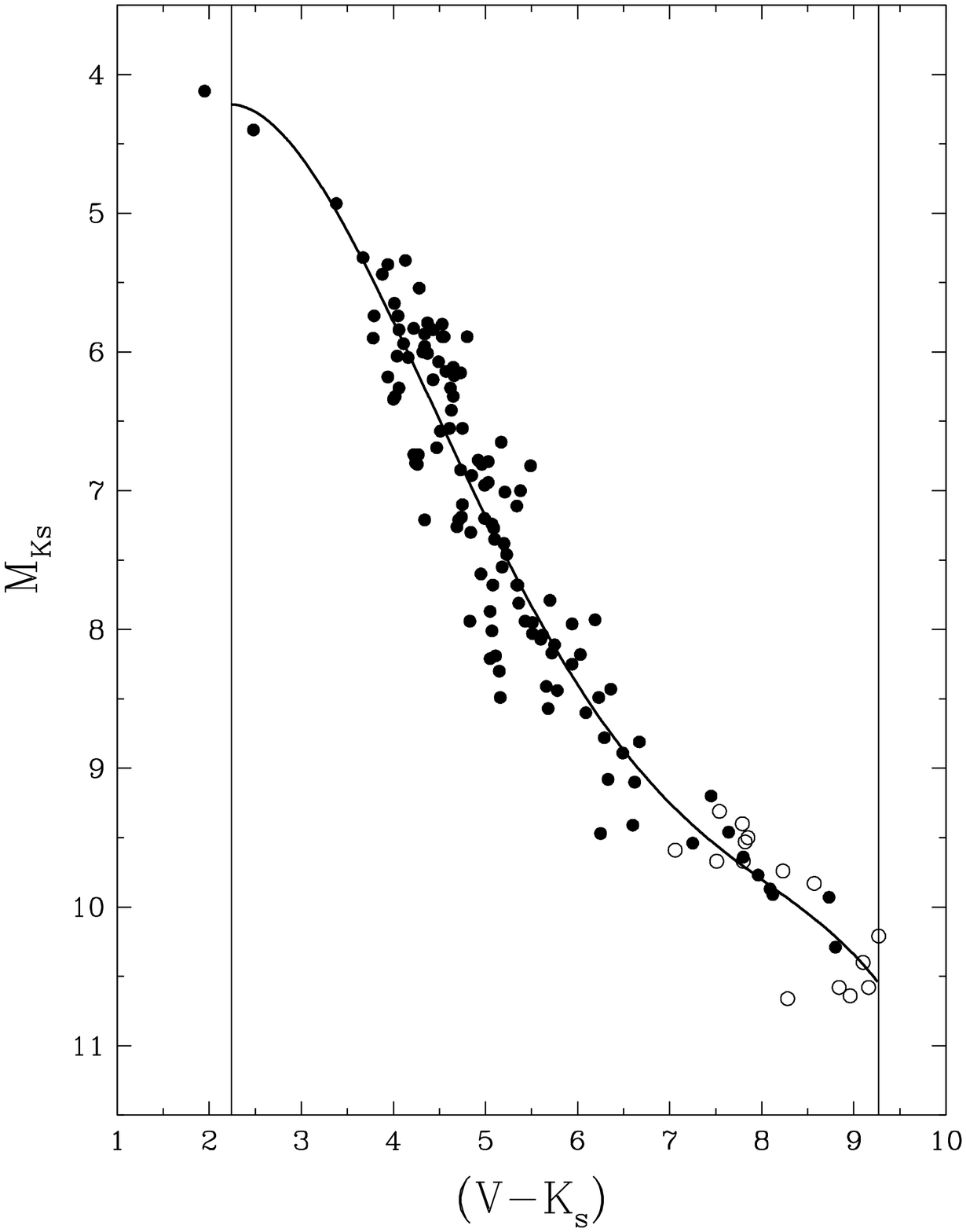}
\label{fig.fitone.wd}
\end{figure}

\begin{figure}
\plotone{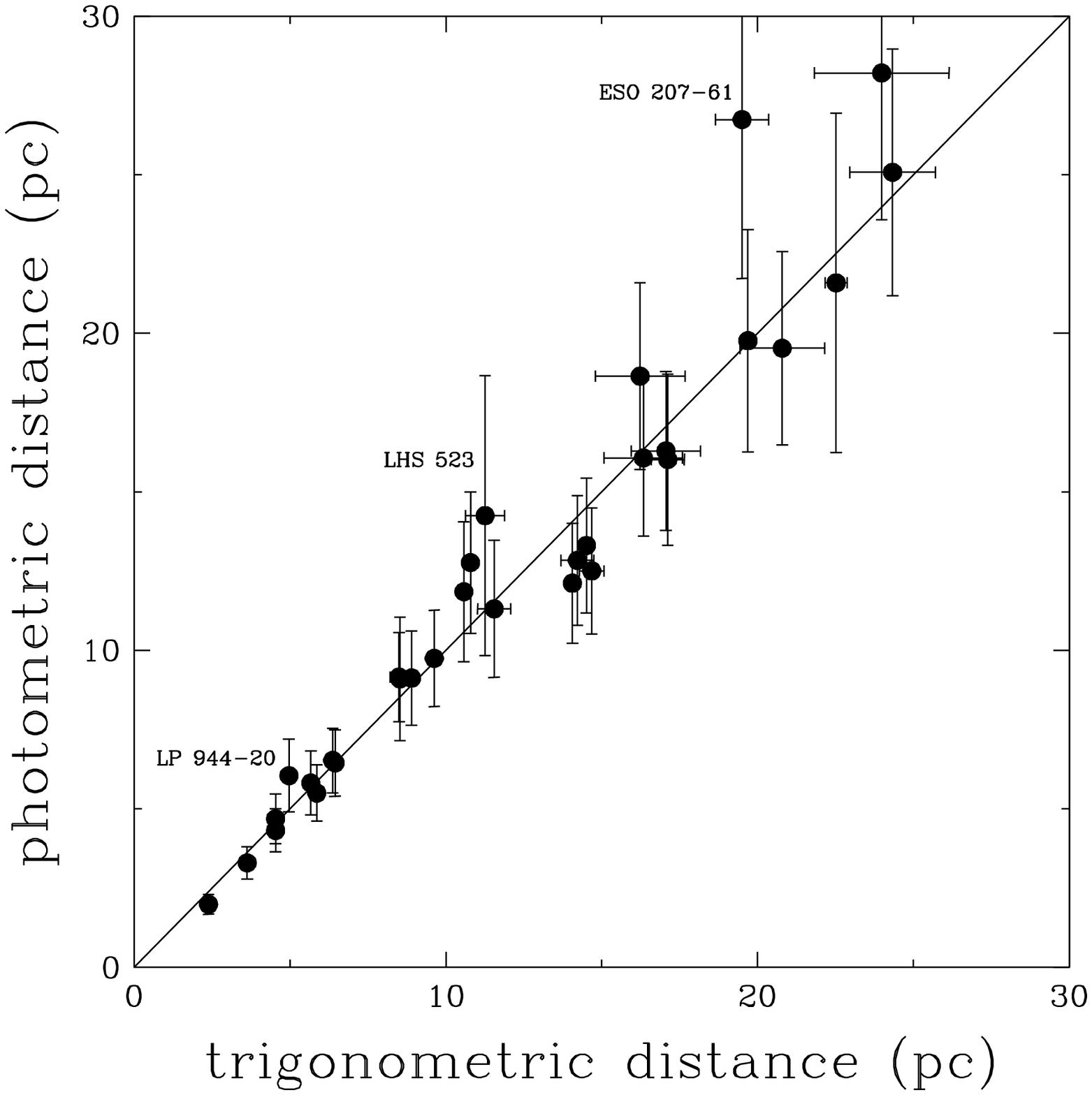}
\label{fig.distances}
\end{figure}

%%%%%%%%%%%%%%%%%%%%%%%%%%%%%% TABLE1: SCR AND CTIO OBSERVATIONS %%%%%%%%%%%%%

\voffset100pt{
{
\begin{deluxetable}{lrcccrrrcrrrlcl}
\rotate
\tabletypesize{\scriptsize}
\tablecaption{Trigonometric parallaxes, VRIJHK$_s$ photometry, and
spectral types for SCR stars, recently discovered nearby red dwarfs,
and stars with $\mu$ $\ge$ 1.0\arcsec/yr recovered in the SCR
search. \label{tbl-1}}
\tablewidth{0pt}

\tablehead{\vspace{-00pt} \\
           \colhead{Name}&
           \colhead{LHS}&
           \colhead{RA (J2000.0) DEC}&
           \colhead{Mean $\pi$$_{trig}$}&
           \colhead{Ref}&
           \colhead{V$_J$}&
           \colhead{R$_C$}&
           \colhead{I$_C$}&
           \colhead{Ref}&
           \colhead{J}&
           \colhead{H}&
           \colhead{K$_s$}&
           \colhead{SpType}&
           \colhead{Ref}&
           \colhead{Notes}}
\startdata
\vspace{-00pt} \\
\tableline \vspace{-00pt} \\
\multicolumn{15}{c}{New SCR Discoveries} \\
\tableline \vspace{-00pt} \\

SCR 0342-6407   &  \nodata &  03 42 57.4 $-$64 07 56 &       \nodata & \nodata & 15.99 & 14.62 & 12.92 &   1   & 11.32 & 10.89 & 10.58 & M4.5V &   1   &  \tablenotemark{a}   \\
SCR 0420-7006   &  \nodata &  04 20 12.6 $-$70 05 59 &       \nodata & \nodata & 17.17 & 15.56 & 13.49 &   1   & 11.19 & 10.59 & 10.25 & M6.0V &   1   &                      \\
SCR 0630-7643AB &  \nodata &  06 30 46.6 $-$76 43 09 &       \nodata & \nodata & 14.91 & 13.32 & 11.24 &   1   &  8.89 &  8.28 &  7.92 & M6.0VJ&   1   &  \tablenotemark{b}   \\
SCR 0702-6102   &  \nodata &  07 02 50.3 $-$61 02 48 &       \nodata & \nodata & 16.61 & 14.75 & 12.48 &   1   & 10.36 &  9.85 &  9.52 & M6.5V &   1   &                      \\
SCR 0723-8015   &  \nodata &  07 23 59.7 $-$80 15 18 &       \nodata & \nodata & 17.41 & 15.60 & 13.41 &   1   & 11.30 & 10.82 & 10.44 & M6.0V &   1   &                      \\
SCR 1138-7721   &  \nodata &  11 38 16.8 $-$77 21 49 &       \nodata & \nodata & 14.78 & 13.20 & 11.25 &   1   &  9.40 &  8.89 &  8.52 & M5.0V &   1   &  \tablenotemark{a,c} \\
SCR 1845-6357   &  \nodata &  18 45 05.3 $-$63 57 48 &       \nodata & \nodata & 17.40 & 15.00 & 12.47 &   1   &  9.54 &  8.97 &  8.51 & M8.5V &   1   &  \tablenotemark{a,d} \\
SCR 1848-6855   &  \nodata &  18 48 21.0 $-$68 55 34 &       \nodata & \nodata & 16.87 & 15.67 & 13.82 &   1   & 11.89 & 11.40 & 11.10 & M5.0V &   1   &  \tablenotemark{a}   \\
SCR 2012-5956   &  \nodata &  20 12 31.8 $-$59 56 52 &       \nodata & \nodata & 15.82 & 15.39 & 15.00 &   1   & 14.93 & 15.23 & 15.41 & DC/DQ &   1   &  \tablenotemark{a,e} \\
\tableline \vspace{-00pt} \\			      	  	      	 	 			                         		  					       
\multicolumn{15}{c}{Recently Discovered Nearby Late-Type M Dwarfs} \\      				         	       			  				       
\tableline \vspace{-00pt} \\			      	  	      	 	 			                         		  					       
SO 0253+1652    &  \nodata &  02 53 00.9 $+$16 52 53 &       \nodata & \nodata & 15.20 & 13.30 & 10.96 &   1   &  8.39 &  7.88 &  7.59 & M7.0V &   1   &  \tablenotemark{f} \\
LP 775-31       &  \nodata &  04 35 16.1 $-$16 06 57 &       \nodata & \nodata & 17.82 & 15.78 & 13.36 &   1   & 10.41 &  9.78 &  9.35 & M7.0V &   1   &  \tablenotemark{g} \\
LP 655-48       &  \nodata &  04 40 23.3 $-$05 30 08 &       \nodata & \nodata & 17.86 & 15.95 & 13.63 &   1   & 10.66 &  9.99 &  9.55 & M7.0V &   1   &  \tablenotemark{h} \\
LHS 2021        &    2021  &  08 30 32.6 $+$09 47 16 &       \nodata & \nodata & 19.38 & 17.22 & 14.97 &   1   & 11.89 & 11.17 & 10.76 & M7.5V &   1   &  \tablenotemark{i} \\
LHS 2090        &    2090  &  09 00 23.6 $+$21 50 05 &       \nodata & \nodata & 16.10 & 14.09 & 11.85 &   1   &  9.44 &  8.84 &  8.44 & M6.0V &   1   &  \tablenotemark{j} \\
DEN 1048-3956   &  \nodata &  10 48 14.6 $-$39 56 07 &       \nodata & \nodata & 17.33 & 14.97 & 12.46 &   1   &  9.54 &  8.91 &  8.45 & M8.5V &   1   &  \tablenotemark{k} \\
LHS 325a        &     325a &  12 23 56.2 $-$27 57 46 &       \nodata & \nodata & 18.39 & 16.42 & 14.20 &   1   & 11.98 & 11.40 & 11.07 & M6.0V &   1   &  \tablenotemark{l} \\
LSR 1826+3014   &  \nodata &  18 26 11.0 $+$30 14 19 &       \nodata & \nodata & 19.36 & 17.40 & 14.35 &   2   & 11.66 & 11.18 & 10.81 & M8.5V &   2   &  \tablenotemark{m} \\
\tableline \vspace{-00pt} \\			      	  		 	 	 		         	                	  					       
\multicolumn{15}{c}{Stars with $\mu$ $\ge$ 1.0\arcsec/yr in Region DEC $-$57.5 to $-$90 Fainter than $m_R$ = 10} \\
\tableline \vspace{-00pt} \\			      	  		 	 	 		         	                	  					       
GJ 1022         &     124  &  00 49 29.1 $-$61 02 33 &       \nodata & \nodata & 12.18 & 11.19 &  9.93 &   1   &  8.63 &  8.09 &  7.84 & M2.5V &   3   &                    \\
GJ   45         &     128  &  00 57 19.7 $-$62 14 44 & .05175 .00106 &   4,5   &  9.47 &  8.68 &  7.98 &   6   &  7.08 &  6.49 &  6.28 & K5.0V &   3   &                    \\
WD 0141-675     &     145  &  01 43 01.0 $-$67 18 30 &       \nodata & \nodata & 13.83 & 13.54 & 13.24 &   1   & 12.87 & 12.66 & 12.58 & DA    &   1   &  \tablenotemark{n} \\
GJ   85         &     150  &  02 07 23.3 $-$66 34 12 & .06460 .01780 &    4    & 11.50 & 10.49 &  9.32 &   6   &  8.13 &  7.61 &  7.36 & M1.5V &   3   &                    \\
GJ  118         &     160  &  02 52 22.2 $-$63 40 48 & .08682 .00188 &   4,5   & 11.38 & 10.32 &  8.99 &   6   &  7.67 &  7.12 &  6.83 & M2.5V &   3   &                    \\
GJ  181.1       &     199  &  04 55 58.0 $-$61 09 47 & .04440 .01080 &    4    & 12.05 & 11.14 & 10.15 &   6   &  9.04 &  8.51 &  8.31 & K7.0V &   3   &                    \\
GJ 1077         &     205  &  05 17 00.0 $-$78 17 20 & .07750 .01100 &    4    & 11.90 & 10.87 &  9.49 &   7   &  8.07 &  7.44 &  7.20 & M2.0V &   3   &                    \\
GJ  293         &      34  &  07 53 08.2 $-$67 47 32 & .14120 .00840 &    4    & 13.60 & 13.21 & 12.85 &   6   & 12.73 & 12.48 & 12.36 & DQ9   &   8   &  \tablenotemark{o} \\
GJ 1123         &     263  &  09 17 05.3 $-$77 49 23 &       \nodata & \nodata & 13.14 & 11.80 & 10.10 &   1   &  8.33 &  7.77 &  7.45 & M4.5V &   9   &                    \\
GJ  345         &     268  &  09 24 21.0 $-$80 31 21 & .01653 .00098 &   4,5   & 10.15 &  9.80 &  9.42 &   6   &  8.89 &  8.53 &  8.46 & F-G   &  10   &                    \\
GJ 1128         &     271  &  09 42 46.4 $-$68 53 06 &       \nodata & \nodata & 12.72 & 11.35 &  9.61 &   1   &  7.95 &  7.39 &  7.04 & M4.5V &   9   &                    \\
LHS 288         &     288  &  10 44 21.2 $-$61 12 36 & .22250 .01130 &    4    & 13.90 & 12.31 & 10.27 &   1   &  8.49 &  8.05 &  7.73 & M5.5V &   1   &  \tablenotemark{p} \\
GJ  422         &      40  &  11 16 00.2 $-$57 32 52 & .07954 .00267 &   4,5   & 11.64 & 10.57 &  9.17 &   6   &  7.81 &  7.30 &  7.04 & M3.5V &   3   &  \tablenotemark{q} \\
GJ  440         &      43  &  11 45 42.9 $-$64 50 30 & .21657 .00201 &   4,5   & 11.47 & 11.30 & 11.15 &   1   & 11.19 & 11.13 & 11.10 & DC/DQ &   1   &  \tablenotemark{r} \\
GJ  467 A       &     328  &  12 28 40.0 $-$71 27 52 & .02150 .01910 &    4    & 13.65 & 12.55 & 11.15 &   6   &  9.81 &  9.30 &  9.05 & M3.0V &   1   &                    \\
GJ  467 B       &     329  &  12 28 43.1 $-$71 27 57 & .02150 .01910 &    4    & 15.75 & 14.38 & 12.66 &   6   & 10.98 & 10.50 & 10.18 & M4.5V &   1   &                    \\
GJ  551         &      49  &  14 29 43.0 $-$62 40 47 & .76876 .00030 &  4,5,11 & 11.09 &  9.42 &  7.37 &   1   &  5.36 &  4.84 &  4.38 & M5.5V &  12   &  \tablenotemark{s} \\
LHS 475         &     475  &  19 20 54.3 $-$82 33 16 &       \nodata & \nodata & 12.68 & 11.50 & 10.00 &   1   &  8.56 &  8.00 &  7.69 & M3.0V &   3   &                    \\
GJ 1251         &     493  &  20 28 03.7 $-$76 40 16 &       \nodata & \nodata & 13.83 & 12.67 & 11.11 &   1   &  9.36 &  8.88 &  8.60 & M4.5V &   3   &                    \\
GJ  808         &     499  &  20 51 41.6 $-$79 18 40 & .06300 .01170 &    4    & 11.83 & 10.83 &  9.65 &   6   &  8.46 &  7.91 &  7.66 & M1.5V &   3   &                    \\
PJH 2115-7541   &  \nodata &  21 15 15.1 $-$75 41 52 &       \nodata & \nodata & 14.46 & 13.24 & 11.66 &   1   & 10.14 &  9.60 &  9.33 & M3.0V &   1   &  \tablenotemark{t} \\
SSPM J2231-7515 &  \nodata &  22 30 33.6 $-$75 15 24 &       \nodata & \nodata & 16.90 & 16.19 & 15.56 &   1   & 14.86 & 14.82 & 14.72 & DX14  &   8   &  \tablenotemark{u} \\
SSPM J2231-7514 &  \nodata &  22 30 40.0 $-$75 13 55 &       \nodata & \nodata & 16.59 & 15.95 & 15.36 &   1   & 14.66 & 14.66 & 14.44 & DX12  &   8   &  \tablenotemark{v} \\
GJ  877         &     531  &  22 55 45.5 $-$75 27 31 & .11610 .00132 &   4,5   & 10.38 &  9.31 &  7.95 &   6   &  6.62 &  6.08 &  5.81 & M2.5V &   1   &  \tablenotemark{w} \\
GJ 1277         &     532  &  22 56 24.7 $-$60 03 49 &       \nodata & \nodata & 14.01 & 12.60 & 10.81 &  13   &  8.98 &  8.36 &  8.11 & M5.0V &   1   &  \tablenotemark{x} \\
		    	     		     	    		     		   	            	       	    	       	      		 	    	     	            	  
\enddata			     

\tablenotetext{a}{also in Hambly et al.~2004}
\tablenotetext{b}{binary with separation 1.0\arcsec, J indicates a joint spectrum}
\tablenotetext{c}{found twice in SCR search}
\tablenotetext{d}{possibly variable in spectral type}
\tablenotetext{e}{white dwarf; $K_s$ unreliable}
\tablenotetext{f}{M6.5V in Teegarden et al.~2003}
\tablenotetext{g}{M8.0V in McCaughrean et al.~2002; M6.0V in Cruz \& Reid 2002}
\tablenotetext{h}{M7.5V in McCaughrean et al.~2002; M6.0V in Cruz \& Reid 2002}
\tablenotetext{i}{error in V = 0.4 mag}
\tablenotetext{j}{M6.5V in Scholz et al.~2001}
\tablenotetext{k}{M9.0V in Delfosse et al.~2001; M8.0V in Gizis 2002; possibly variable in spectral type}
\tablenotetext{l}{M6.0V in Bessell 1991 (listed as LHS 325 instead of LHS 325a)}
\tablenotetext{m}{northern target not observed by RECONS from Chile}
\tablenotetext{n}{white dwarf; DA7 in McCook \& Sion 2003}
\tablenotetext{o}{white dwarf}
\tablenotetext{p}{missed in SCR search; $VRI$ = 13.92, 12.33, 10.31 in Bessell 1991; blended in 2MASS}
\tablenotetext{q}{missed in SCR search}
\tablenotetext{r}{white dwarf; missed in SCR search; DQ6 in McCook \& Sion 2003}
\tablenotetext{s}{Proxima; missed in SCR search; $VRI$ = 11.05, 9.43, 7.43 in Bessell 1990}
\tablenotetext{t}{also SCR 2115-7541}
\tablenotetext{u}{also SCR 2230-7515; white dwarf; $V$ = 16.87 in Scholz et al.~2002}
\tablenotetext{v}{also SCR 2230-7513; white dwarf; $V$ = 16.60 in Scholz et al.~2002}
\tablenotetext{w}{M2.5V in Hawley et al.~1996}
\tablenotetext{x}{M4.5V in Hawley et al.~1996}

\tablenotetext{~}{References.---
(1) This paper. 
(2) Lepine et al.~2002. 
(3) Hawley et al.~1996. 
(4) van Altena et al.~1995.
(5) ESA 1997. 
(6) Bessell 1990. 
(7) Leggett 1992. 
(8) McCook \& Sion 2003.
(9) Henry et al.~2002. 
(10) Gliese 1969.
(11) Benedict et al.~1999. 
(12) Henry et al.~1997. 
(13) Patterson et al.~1998.
}

\end{deluxetable}
}}
\clearpage

%%%%%%%%%%%%%%%%%%%%%%%%%% TABLE2: LATE M DWARFS WITHIN 25 PC %%%%%%%%%%%%%%%%

\voffset100pt{
{
\begin{deluxetable}{lrcccrrrcrrrlcl}
\rotate
\tabletypesize{\scriptsize}
\tablecaption{Trigonometric parallaxes, VRIJHK$_s$ photometry, and spectral types for the supplemental sample of late-type M dwarfs. \label{tbl-2}}
\tablewidth{0pt}

\tablehead{\vspace{-00pt} \\
           \colhead{Name}&
           \colhead{LHS}&
           \colhead{RA (J2000.0) DEC}&
           \colhead{Mean $\pi$$_{trig}$}&
           \colhead{Ref}&
           \colhead{V$_J$}&
           \colhead{R$_C$}&
           \colhead{I$_C$}&
           \colhead{Ref}&
           \colhead{J}&
           \colhead{H}&
           \colhead{K$_s$}&
           \colhead{SpType}&
           \colhead{Ref}&
           \colhead{Notes}}
\startdata
\vspace{-00pt} \\
\tableline \vspace{-00pt} \\			      	  		 	 	 					                 
\multicolumn{15}{c}{Late-Type M Dwarfs within 25 Parsecs} \\		 	 	 					                 
\tableline \vspace{-00pt} \\			      	  		 	 	 		

BRI 0021-0214   &  \nodata &  00 24 24.6  $-$01 58 20 & .08660 .00400 & 1        &\nodata &  17.42 &  15.16 & 2     &  11.99 &  11.08 &  10.54 & M9.0V &   3   & \tablenotemark{a} \\
RG 0050-2722    &  \nodata &  00 52 54.7  $-$27 06 00 & .04171 .00372 & 4,5      &  21.50 &\nodata &  16.82 & 5,6   &  13.61 &  12.98 &  12.54 & M8.0V &   7   &                   \\
2MA 0149+2956   &  \nodata &  01 49 09.0  $+$29 56 12 & .04440 .00070 & 2        &  21.25 &  18.94 &  16.81 & 2     &  13.45 &  12.58 &  11.98 & M9.5V &   8   &                   \\
LHS 1375        &    1375  &  02 16 29.9  $+$13 35 13 & .11770 .00400 & 4        &  15.79 &\nodata &\nodata & 9     &   9.87 &   9.31 &   8.98 & M6.0V &   3   & \tablenotemark{b} \\
LP 771-21       &  \nodata &  02 48 41.0  $-$16 51 22 & .06160 .00543 & 5        &\nodata &\nodata &  15.42 & 5     &  12.55 &  11.87 &  11.42 & M8.0V &  10   &                   \\
LP 412-31       &  \nodata &  03 20 59.7  $+$18 54 23 & .06890 .00060 & 2        &  19.21 &  16.98 &  14.70 & 2     &  11.76 &  11.07 &  10.64 & M8.0V &  11   & \tablenotemark{c} \\
LP 944-20       &  \nodata &  03 39 35.3  $-$35 25 44 & .20140 .00421 & 5        &\nodata &\nodata &  14.16 & 5     &  10.73 &  10.02 &   9.55 & M9.0V &   3   & \tablenotemark{d} \\
LHS 1604        &    1604  &  03 51 00.0  $-$00 52 45 & .06810 .00180 & 4        &  18.02 &\nodata &  13.80 & 12    &  11.30 &  10.61 &  10.23 & M6.0V &  13   & \tablenotemark{e} \\
LHS 191         &     191  &  04 26 19.9  $+$03 36 36 & .05840 .00180 & 4        &  18.51 &  16.24 &  13.96 & 14    &  11.62 &  11.07 &  10.69 & M6.5V &  15   &                   \\
ESO 207-61      &  \nodata &  07 07 53.3  $-$49 00 50 & .05129 .00226 & 5,16     &  20.39 &  18.63 &  16.23 & 17    &  13.23 &  12.54 &  12.11 & M9.0V:&  18   & \tablenotemark{f} \\
GJ 283B         &     234  &  07 40 19.4  $-$17 24 46 & .11240 .00270 & 4        &  16.54 &  14.68 &  12.43 & 19    &  10.16 &   9.63 &   9.29 & M6.5V &   3   & \tablenotemark{g} \\
GJ 1111         &     248  &  08 29 49.3  $+$26 46 34 & .27580 .00300 & 4        &  14.90 &  12.90 &  10.64 & 19    &   8.24 &   7.62 &   7.26 & M6.5V &   3   & \tablenotemark{h} \\
LHS 2026        &    2026  &  08 32 30.5  $-$01 34 39 & .05080 .00050 & 4        &  18.94 &  16.69 &  14.32 & 14    &  12.04 &  11.48 &  11.14 & M6.0V &  14   &                   \\
GJ 316.1        &    2034  &  08 40 29.7  $+$18 24 09 & .07110 .00100 & 4        &  17.59 &\nodata &  13.45 & 20    &  11.05 &  10.42 &  10.05 & M6.0V &  13   & \tablenotemark{i} \\
LHS 2065        &    2065  &  08 53 36.2  $-$03 29 32 & .11730 .00150 & 4        &  18.74 &  16.74 &  14.54 & 14    &  11.21 &  10.47 &   9.94 & M9.0V &   3   & \tablenotemark{j} \\
LHS 292         &     292  &  10 48 12.6  $-$11 20 10 & .22030 .00360 & 4        &  15.73 &  13.67 &  11.33 & 14    &   8.86 &   8.26 &   7.93 & M7.0V &  21   & \tablenotemark{k} \\
LHS 2314        &    2314  &  10 49 03.4  $+$05 02 23 & .04110 .00230 & 4        &  19.11 &\nodata &  14.91 & 12    &  12.54 &  11.97 &  11.60 & M6.0V &  13   &                   \\
GJ 406          &      36  &  10 56 28.9  $+$07 00 53 & .41910 .00210 & 4        &  13.53 &  11.67 &   9.50 & 19    &   7.09 &   6.48 &   6.08 & M6.0V &   3   & \tablenotemark{l} \\
LHS 2351        &    2351  &  11 06 19.0  $+$04 28 33 & .04810 .00314 & 5        &  19.56 &  17.25 &  14.91 & 14    &  12.33 &  11.72 &  11.33 & M6.5V &  14   &                   \\
LHS 2471        &    2471  &  11 53 52.7  $+$06 59 56 & .07030 .00260 & 4        &  18.11 &\nodata &  13.66 & 14    &  11.26 &  10.66 &  10.26 & M6.0V &  13   & \tablenotemark{m} \\
BRI 1222-1222   &  \nodata &  12 24 52.2  $-$12 38 36 & .05860 .00380 & 5        &\nodata &\nodata &  15.74 & 5     &  12.57 &  11.82 &  11.35 & M9.0V &  11   &                   \\
LHS 2924        &    2924  &  14 28 43.2  $+$33 10 39 & .09267 .00128 & 1,4      &  19.58 &\nodata &  15.21 & 20    &  11.99 &  11.23 &  10.74 & M9.0V &  11   &                   \\
LHS 2930        &    2930  &  14 30 37.8  $+$59 43 25 & .10380 .00130 & 4        &  17.88 &\nodata &  13.31 & 20    &  10.79 &  10.14 &   9.79 & M6.5V &   7   &                   \\
LHS 3003        &    3003  &  14 56 38.3  $-$28 09 49 & .15705 .00259 & 4,5      &  17.05 &  14.88 &  12.53 & 14    &   9.97 &   9.32 &   8.93 & M7.0V &   3   & \tablenotemark{n} \\
TVLM 513-46546  &  \nodata &  15 01 08.2  $+$22 50 02 & .09450 .00060 & 1,2      &  19.87 &  17.53 &  15.16 & 2     &  11.87 &  11.18 &  10.71 & M9.0V &   3   & \tablenotemark{o} \\
TVLM 868-110639 &  \nodata &  15 10 16.8  $-$02 41 08 & .06120 .00470 & 1        &\nodata &\nodata &  15.79 & 1     &  12.61 &  11.84 &  11.35 & M9.0V &  11   &                   \\
GJ 644C         &     429  &  16 55 35.3  $-$08 23 41 & .15497 .00056 & 4,22,23  &  16.78 &  14.60 &  12.18 & 19    &   9.78 &   9.20 &   8.82 & M7.0V &   3   & \tablenotemark{p} \\
2MA 1835+3259   &  \nodata &  18 35 37.9  $+$32 59 53 & .17650 .00050 & 24       &  18.27 &\nodata &  13.46 & 24    &  10.27 &   9.62 &   9.17 & M8.5V &  24   &                   \\
GJ 752B         &     474  &  19 16 57.6  $+$05 09 02 & .17079 .00061 & 1,4,5,22 &  17.50 &  15.10 &  12.84 & 14,20 &   9.91 &   9.23 &   8.77 & M8.0V &   3   & \tablenotemark{q} \\
GJ 1245B        &    3495  &  19 53 55.2  $+$44 24 54 & .22020 .00100 & 4        &  14.01 &  12.36 &  10.27 & 25    &   8.28 &   7.73 &   7.39 & M6.0V &  15   &                   \\
LHS 523         &     523  &  22 28 54.4  $-$13 25 19 & .08880 .00490 & 4        &  16.90 &  14.90 &  12.56 & 19    &  10.77 &  10.22 &   9.84 & M6.5V &  15   & \tablenotemark{r} \\

\enddata 

\tablenotetext{a}{parallax in van Altena et al.~1995 is superseded by Tinney et al.~1995 value; M9.5V: in Kirkpatrick et al.~1995; M9.5V in Kirkpatrick et al.~2000}
\tablenotetext{b}{M5.5V in Reid et al.~1995}
\tablenotetext{c}{M9.0V in Gizis et al.~2000; M9.0V in Cruz \& Reid 2002}
\tablenotetext{d}{M9.0V in Kirkpatrick et al.~2000; M9.5V in McCaughrean et al.~2002; M9.5V in Reid \& Cruz 2002}
\tablenotetext{e}{M6.0V in Cruz \& Reid 2002}
\tablenotetext{f}{parallax in van Altena et al.~1995 is superseded by Ianna \& Fredrick 1995 value; photometry is quoted as being ``Kron-Cousins'' but filter set likely matches Cousins; 
                  spectral type determined by comparison of printed published spectrum to standards by our group}
\tablenotetext{g}{M6.0V in Henry et al.~1994}
\tablenotetext{h}{M6.5V in Kirkpatrick et al.~1991}
\tablenotetext{i}{M6.0V in Bessell 1991}
\tablenotetext{j}{M9.0V in Kirkpatrick et al.~1991}
\tablenotetext{k}{M6.5V in Henry et al.~1994}
\tablenotetext{l}{M6.0V in Kirkpatrick et al.~1991}
\tablenotetext{m}{M6.5V in Bessell 1991}
\tablenotetext{n}{M6.5V in Bessell 1991; M7.0V in Kirkpatrick et al.~1995}
\tablenotetext{o}{M8.5V in Kirkpatrick et al.~1995}
\tablenotetext{p}{VB 8; several parallaxes for GJ 644ABD and GJ 643, members of the same system, included; M7.0V in Kirkpatrick et al.~1991}
\tablenotetext{q}{VB 10; several parallaxes for GJ 752A included; parallax in van Altena et al.~1995 was recalculated without value superseded by Tinney et al.~1995; M8.0V in Kirkpatrick et al.~1991}
\tablenotetext{r}{M6.0V in Bessell 1991; M6.5V in Cruz \& Reid 2002}

\tablenotetext{~}{References.---
(1) Tinney et al.~1995.
(2) Dahn et al.~2002.
(3) This paper.
(4) van Altena et al.~1995.
(5) Tinney 1996.
(6) Reid \& Gilmore 1984.
(7) Kirkpatrick et al.~1993.
(8) Kirkpatrick et al.~1999.
(9) Dahn et al.~1988. 
(10) Kirkpatrick et al.~1997.
(11) Kirkpatrick et al.~1995.
(12) Monet et al.~1992. 
(13) Reid et al.~1995. 
(14) Bessell 1991. 
(15) Kirkpatrick et al.~1991.
(16) Ianna \& Fredrick 1995.
(17) Ruiz et al.~1991.
(18) Ruiz et al.~1995.
(19) Bessell 1990.
(20) Leggett 1992.
(21) Henry et al.~1994.
(22) ESA 1997.
(23) Soderhjelm 1999.
(24) Reid et al.~2003.
(25) Weis 1996.
}

\end{deluxetable}
}}
\clearpage

%%%%%%%%%%%%%%%%%%%%%%%%%%% TABLE3: DISTANCE RELATIONS %%%%%%%%%%%%%%%%%%%%%%%

\hoffset-040pt{
\begin{deluxetable}{ccccrrrrrc}
\tabletypesize{\scriptsize}
\tablecaption{Details for photometric distance relations. \label{relations}}
\tablewidth{0pt}
\tablehead{\vspace{-00pt} \\
           \colhead{}               &
           \colhead{Applicable}     &
           \colhead{\# RECONS}      &
           \colhead{\# Very Red}    &
           \colhead{Coeff \#1}      &
           \colhead{Coeff \#2}      &
           \colhead{Coeff \#3}      &
           \colhead{Coeff \#4}      &
           \colhead{Coeff \#5}      &
           \colhead{RMS} \\

           \colhead{Color}          &
           \colhead{Range}          &
           \colhead{Stars}          &
           \colhead{Stars}          &
           \colhead{$\times$ (color)$^4$}&
           \colhead{$\times$ (color)$^3$}&
           \colhead{$\times$ (color)$^2$}&
           \colhead{$\times$ (color)}    &
           \colhead{(constant)}          &
           \colhead{(mag)}}
\startdata
$(V-R)$ &  0.53 to 2.40 &  117 &   8 &  $+$  2.79703  & $-$ 17.48617  & $+$ 36.67711  & $-$ 25.90589 & $+$  9.96960  &  0.40  \\
$(V-I)$ &  0.88 to 4.81 &  119 &  15 &  $+$  0.02853  & $-$  0.49504  & $+$  2.64479  & $-$  3.51296 & $+$  5.62135  &  0.40  \\
$(V-J)$ &  2.51 to 8.00 &  115 &  15 &  $+$  0.02447  & $-$  0.52310  & $+$  3.91317  & $-$ 10.94674 & $+$ 15.31851  &  0.39  \\
$(V-H)$ &  3.59 to 8.69 &  100 &  15 &  $+$  0.03207  & $-$  0.77797  & $+$  6.74382  & $-$ 23.61879 & $+$ 34.23360  &  0.42  \\
$(V-K)$ &  2.24 to 9.27 &  119 &  15 &  $+$  0.00959  & $-$  0.23953  & $+$  2.05071  & $-$  5.98231 & $+$  9.77683  &  0.42  \\
$(R-I)$ &  0.43 to 2.42 &  117 &   9 &  $-$  1.08390  & $+$  5.68997  & $-$  9.78999  & $+$  9.22596 & $+$  1.54462  &  0.41  \\ 
$(R-J)$ &  1.64 to 5.66 &  114 &   9 &  $+$  0.07380  & $-$  1.15011  & $+$  6.26647  & $-$ 12.52051 & $+$ 13.44932  &  0.41  \\
$(R-H)$ &  2.68 to 6.36 &   99 &   9 &  $+$  0.10427  & $-$  1.91432  & $+$ 12.58352  & $-$ 33.56316 & $+$ 36.76955  &  0.45  \\
$(R-K)$ &  1.63 to 6.97 &  117 &   9 &  $+$  0.01785  & $-$  0.37226  & $+$  2.59680  & $-$  5.75029 & $+$  8.19804  &  0.45  \\
$(I-J)$ &  0.88 to 3.36 &  116 &  19 &  $+$  0.58092  & $-$  4.69507  & $+$ 12.35365  & $-$  9.20851 & $+$  6.22309  &  0.45  \\
$(I-H)$ &  1.67 to 4.23 &  101 &  19 &  $+$  0.14094  & $-$  1.31052  & $+$  3.12906  & $+$  2.68748 & $-$  2.62035  &  0.54  \\
$(I-K)$ &  1.07 to 4.83 &  120 &  19 &  $+$  0.19771  & $-$  2.44679  & $+$ 10.18426  & $-$ 14.30638 & $+$ 10.38741  &  0.52  \\
\enddata													  
\end{deluxetable}												  
}
\clearpage													  
														  
%%%%%%%%%%%%%%%%%%%%%%%%%% TABLE4: DISTANCE ESTIMATES %%%%%%%%%%%%%%%%%%%%%%%%

\hoffset-040pt{
\begin{deluxetable}{lcrcrrrl}
\tabletypesize{\scriptsize}
\tablecaption{Distance estimates from new photometric parallax relations. \label{distances}}
\tablewidth{0pt}
\tablehead{\vspace{-00pt} \\
           \colhead{Name}           &
           \colhead{Mean $\pi$$_{trig}$}&
           \colhead{M$_{Ks}$}       &
           \colhead{\# colors}      &
           \colhead{Est Dist (pc)}  &
           \colhead{True Dist (pc)} &
           \colhead{\% Diff}        &
           \colhead{Notes}}
\startdata
\vspace{-00pt} \\
\multicolumn{8}{c}{New SCR Discoveries} \\
\tableline \vspace{-00pt} \\
SCR 0342-6407   &       \nodata & \nodata & 12 & 38.08 $\pm$ 7.79 &           \nodata & \nodata & 39.3 $\pm$ 11.7 pc in Hambly et al.~2004     \\
SCR 0420-7006   &       \nodata & \nodata & 12 & 15.41 $\pm$ 2.57 &           \nodata & \nodata &                                              \\
SCR 0630-7643AB &       \nodata & \nodata & 12 &  6.95 $\pm$ 1.21 &           \nodata & \nodata & assuming $\Delta$mag = 0.25 in all filters \\
SCR 0702-6102   &       \nodata & \nodata & 12 & 10.84 $\pm$ 2.06 &           \nodata & \nodata &                                              \\
SCR 0724-8015   &       \nodata & \nodata & 12 & 17.16 $\pm$ 3.10 &           \nodata & \nodata &                                              \\
SCR 1138-7721   &       \nodata & \nodata & 12 &  9.43 $\pm$ 1.68 &           \nodata & \nodata & 8.8 $\pm$ 1.7 pc in Hambly et al.~2004       \\
SCR 1845-6357   &       \nodata & \nodata & 10 &  4.63 $\pm$ 0.75 &           \nodata & \nodata & 3.5 $\pm$ 0.7 pc in Hambly et al.~2004       \\
SCR 1848-6855   &       \nodata & \nodata & 12 & 37.03 $\pm$ 9.43 &           \nodata & \nodata & 34.8 $\pm$ 9.8 pc in Hambly et al.~2004      \\
SCR 2012-5956   &       \nodata & \nodata &  1 & 17.37 $\pm$ 3.47 &           \nodata & \nodata & white dwarf distance estimate                \\
\tableline \vspace{-00pt} \\			      	   	     	      	 	   			                       		 
\multicolumn{8}{c}{Recently Discovered Nearby Late-Type M Dwarfs} \\	     	      		   					       	       	
\tableline \vspace{-00pt} \\			      	   	     	      	 	   			                       		 
SO 0253+1652    &       \nodata & \nodata & 12 &  3.73 $\pm$ 0.59 &           \nodata & \nodata & \tablenotemark{a}                            \\
LP 775-31       &       \nodata & \nodata & 12 &  7.33 $\pm$ 1.19 &           \nodata & \nodata & \tablenotemark{b}                            \\
LP 655-48       &       \nodata & \nodata & 12 &  8.24 $\pm$ 1.40 &           \nodata & \nodata & \tablenotemark{c}                            \\
LHS 2021        &       \nodata & \nodata & 12 & 13.80 $\pm$ 2.34 &           \nodata & \nodata &                                              \\
LHS 2090        &       \nodata & \nodata & 12 &  5.67 $\pm$ 0.88 &           \nodata & \nodata & \tablenotemark{d}                            \\
DEN 1048-3956   &       \nodata & \nodata & 10 &  4.53 $\pm$ 0.73 &           \nodata & \nodata & \tablenotemark{e}                            \\
LHS 325a        &       \nodata & \nodata & 12 & 20.74 $\pm$ 3.43 &           \nodata & \nodata &                                              \\
LSR 1826+3014   &       \nodata & \nodata & 10 & 14.48 $\pm$ 2.52 &           \nodata & \nodata & \tablenotemark{f}                            \\
\tableline \vspace{-00pt} \\			      	   	     	      	 	   			                 			       
\multicolumn{8}{c}{Stars with $\mu$ $\ge$ 1.0\arcsec/yr in Region DEC $-$57.5 to $-$90 Fainter than $m_R$ = 10} \\
\tableline \vspace{-00pt} \\			      	   	     	      	 	   			                 			       
GJ 1022         &       \nodata & \nodata & 12 & 20.49 $\pm$ 3.35 &           \nodata & \nodata & \tablenotemark{g}                            \\
GJ   45         & .05175 .00106 &    4.85 &  7 & 19.18 $\pm$ 3.11 &  19.32 $\pm$ 0.40 & $-$ 0.7 &                                              \\
WD 0141-675     &       \nodata & \nodata &  1 &  9.27 $\pm$ 1.85 &           \nodata & \nodata & white dwarf distance estimate                \\
GJ   85         & .06460 .01780 &    6.42 & 12 & 19.01 $\pm$ 3.00 &  15.48 $\pm$ 4.62 & $+$22.8 & poor parallax                                \\
GJ  118         & .08682 .00188 &    6.52 & 12 & 11.57 $\pm$ 1.80 &  11.52 $\pm$ 0.25 & $+$ 0.5 &                                              \\
GJ  181.1       & .04440 .01080 &    6.55 &  9 & 36.03 $\pm$ 5.65 &  22.52 $\pm$ 5.82 & $+$60.0 & poor parallax                                \\
GJ 1077         & .07750 .01100 &    6.65 & 12 & 12.14 $\pm$ 2.38 &  12.90 $\pm$ 1.87 & $-$ 5.9 & poor parallax                                \\
GJ  293         & .14120 .00840 &   13.11 &  1 &  6.82 $\pm$ 1.36 &   7.08 $\pm$ 0.42 & \nodata & white dwarf distance estimate                \\
GJ 1123         &       \nodata & \nodata & 12 &  7.47 $\pm$ 1.22 &           \nodata & \nodata & \tablenotemark{h}                            \\
GJ  345         & .01653 .00098 &    4.56 &  0 &          \nodata &  60.50 $\pm$ 3.60 & \nodata & too blue for relations                       \\
GJ 1128         &       \nodata & \nodata & 12 &  6.41 $\pm$ 1.01 &           \nodata & \nodata & \tablenotemark{i}                            \\
LHS 288         & .22250 .01130 &    9.46 & 12 &  6.90 $\pm$ 1.74 &   4.49 $\pm$ 0.23 & $+$53.5 & poor parallax                                \\
GJ  422         & .07954 .00267 &    6.54 & 12 & 12.12 $\pm$ 1.92 &  12.57 $\pm$ 0.42 & $-$ 3.6 &                                              \\
GJ  440         & .21657 .00201 &   12.78 &  1 &  4.38 $\pm$ 0.88 &   4.62 $\pm$ 0.04 & \nodata & white dwarf distance estimate                \\
GJ  467 A       & .02150 .01910 &    5.71 & 12 & 30.87 $\pm$ 4.98 &  46.51 $\pm$ huge & $-$33.6 & poor parallax                                \\
GJ  467 B       & .02150 .01910 &    6.84 & 12 & 28.48 $\pm$ 4.79 &  46.51 $\pm$ huge & $-$38.8 & poor parallax                                \\
GJ  551         & .76876 .00030 &    8.81 & 12 &  1.17 $\pm$ 0.19 &   1.30 $\pm$ 0.01 & $-$10.2 & Proxima                                      \\
LHS 475         &       \nodata & \nodata & 12 & 13.07 $\pm$ 2.03 &           \nodata & \nodata &                                              \\
GJ 1251         &       \nodata & \nodata & 12 & 16.05 $\pm$ 3.40 &           \nodata & \nodata &                                              \\
GJ  808         & .06300 .01170 &    6.66 & 12 & 21.41 $\pm$ 7.43 &  15.87 $\pm$ 3.05 & $+$34.9 & poor parallax                                \\
PJH 2115-7541   &       \nodata & \nodata & 12 & 24.84 $\pm$ 3.84 &           \nodata & \nodata &                                              \\
SSSPM J2231-7515&       \nodata & \nodata &  1 & 14.89 $\pm$ 2.98 &           \nodata & \nodata & white dwarf distance estimate                \\
SSSPM J2231-7514&       \nodata & \nodata &  1 & 14.82 $\pm$ 2.96 &           \nodata & \nodata & white dwarf distance estimate                \\
GJ  877         & .11610 .00132 &    6.14 & 12 &  7.10 $\pm$ 1.10 &   8.61 $\pm$ 0.10 & $-$17.5 &                                              \\
GJ 1277         &       \nodata & \nodata & 12 &  8.89 $\pm$ 1.42 &           \nodata & \nodata &                                              \\
\tableline \vspace{-00pt} \\			      	   	  	    	 					                 		   
\multicolumn{8}{c}{Late-Type M Dwarfs within 25 Parsecs} \\	  	 	    	 				                 		   
\tableline \vspace{-00pt} \\			      	   	  	    	 					                 		   
BRI 0021-0214   & .08660 .00400 &   10.23 &  7 & 11.31 $\pm$ 2.16 &  11.55 $\pm$ 0.53 & $-$ 2.1 & \hskip130pt {}                               \\
RG 0050-2722    & .04171 .00372 &   10.64 &  7 & 28.21 $\pm$ 4.63 &  23.98 $\pm$ 2.16 & $+$17.7 &                                              \\
2MA 0149+2956   & .04440 .00070 &   10.21 & 12 & 21.59 $\pm$ 5.35 &  22.52 $\pm$ 0.36 & $-$ 4.2 &                                              \\
LHS 1375        & .11770 .00400 &    9.33 &  3 &  9.16 $\pm$ 1.41 &   8.50 $\pm$ 0.29 & $+$ 7.8 &                                              \\
LP 771-21       & .06160 .00543 &   10.37 &  3 & 18.65 $\pm$ 2.94 &  16.23 $\pm$ 1.44 & $+$14.9 &                                              \\
LP 412-31       & .06890 .00060 &    9.83 & 12 & 13.31 $\pm$ 2.13 &  14.51 $\pm$ 0.13 & $-$ 8.3 & \tablenotemark{j}                            \\
LP 944-20       & .20140 .00421 &   11.07 &  3 &  6.05 $\pm$ 1.15 &   4.97 $\pm$ 0.10 & $+$21.9 &                                              \\
LHS 1604        & .06810 .00180 &    9.40 &  7 & 12.51 $\pm$ 1.99 &  14.68 $\pm$ 0.39 & $-$14.8 & \tablenotemark{k}                            \\
LHS 191         & .05840 .00180 &    9.53 & 12 & 16.02 $\pm$ 2.70 &  17.12 $\pm$ 0.53 & $-$ 6.5 & \hskip130pt {}                               \\
ESO 207-61      & .05129 .00226 &   10.66 & 12 & 26.74 $\pm$ 5.01 &  19.50 $\pm$ 0.86 & $+$37.1 &                                              \\
GJ 283B         & .11240 .00270 &    9.54 & 12 &  9.13 $\pm$ 1.49 &   8.90 $\pm$ 0.21 & $+$ 2.6 &                                              \\
GJ 1111         & .27580 .00300 &    9.46 & 12 &  3.30 $\pm$ 0.51 &   3.63 $\pm$ 0.04 & $-$ 9.1 &                                              \\
LHS 2026        & .05080 .00050 &    9.67 & 12 & 19.77 $\pm$ 3.51 &  19.69 $\pm$ 0.19 & $+$ 0.5 &                                              \\
GJ 316.1        & .07110 .00100 &    9.31 &  7 & 12.12 $\pm$ 1.89 &  14.06 $\pm$ 0.20 & $-$13.9 &                                              \\
LHS 2065        & .11730 .00150 &   10.29 & 12 &  9.10 $\pm$ 1.95 &   8.53 $\pm$ 0.11 & $+$ 6.7 & \tablenotemark{l}                            \\
LHS 292         & .22030 .00360 &    9.64 & 12 &  4.32 $\pm$ 0.68 &   4.54 $\pm$ 0.07 & $-$ 4.9 &                                              \\
LHS 2314        & .04110 .00230 &    9.67 &  7 & 25.08 $\pm$ 3.89 &  24.33 $\pm$ 1.37 & $+$ 3.1 &                                              \\
GJ 406          & .41910 .00210 &    9.20 & 12 &  1.99 $\pm$ 0.31 &   2.39 $\pm$ 0.01 & $-$16.5 &                                              \\
LHS 2351        & .04810 .00314 &    9.74 & 12 & 19.53 $\pm$ 3.05 &  20.79 $\pm$ 1.36 & $-$ 6.1 &                                              \\
LHS 2471        & .07030 .00260 &    9.50 &  7 & 12.84 $\pm$ 2.04 &  14.22 $\pm$ 0.53 & $-$ 9.8 &                                              \\
BRI 1222-1222   & .05860 .00380 &   10.19 &  3 & 16.29 $\pm$ 2.50 &  17.06 $\pm$ 1.11 & $-$ 4.6 &                                              \\
LHS 2924        & .09267 .00128 &   10.58 &  7 & 12.77 $\pm$ 2.23 &  10.79 $\pm$ 0.15 & $+$18.4 &                                              \\
LHS 2930        & .10380 .00130 &    9.87 &  7 &  9.75 $\pm$ 1.52 &   9.63 $\pm$ 0.12 & $+$ 1.2 &                                              \\
LHS 3003        & .15705 .00259 &    9.91 & 12 &  6.53 $\pm$ 1.02 &   6.37 $\pm$ 0.11 & $+$ 2.6 & \tablenotemark{m}                            \\
TVLM 513-46546  & .09450 .00060 &   10.58 & 12 & 11.85 $\pm$ 2.21 &  10.58 $\pm$ 0.07 & $+$12.0 &                                              \\
TVLM 868-110639 & .06120 .00470 &   10.28 &  3 & 16.07 $\pm$ 2.47 &  16.34 $\pm$ 1.26 & $-$ 1.6 &                                              \\
GJ 644C         & .15497 .00056 &    9.77 & 12 &  6.45 $\pm$ 1.05 &   6.45 $\pm$ 0.02 & $+$ 0.0 &                                              \\
2MA 1835+3259   & .17650 .00050 &   10.40 &  7 &  5.82 $\pm$ 1.01 &   5.67 $\pm$ 0.02 & $+$ 2.7 &                                              \\
GJ 752B         & .17079 .00061 &    9.93 & 12 &  5.50 $\pm$ 0.89 &   5.86 $\pm$ 0.02 & $-$ 6.1 &                                              \\
GJ 1245B        & .22020 .00100 &    9.10 & 12 &  4.69 $\pm$ 0.79 &   4.54 $\pm$ 0.02 & $+$ 3.4 &                                              \\
LHS 523         & .08880 .00490 &    9.59 & 12 & 14.25 $\pm$ 4.41 &  11.26 $\pm$ 0.62 & $+$26.5 & \tablenotemark{n}                            \\
\enddata													  

\tablenotetext{a}{2.4 pc $\pm$ 0.5 pc from astrometry and 3.6 $\pm$ 0.4 pc from photometry in Teegarden et al.~2003}
\tablenotetext{b}{6.2--6.5 pc in McCaughrean et al.~2002; 7.4 $\pm$ 1.5 pc in Reid \& Cruz 2002; 11.3 $\pm$ 1.3 pc in Cruz \& Reid 2002}
\tablenotetext{c}{7.9--8.2 pc in McCaughrean et al.~2002; 7.7 $\pm$ 1.5 pc in Reid \& Cruz 2002; 15.3 $\pm$ 2.6 pc in Cruz \& Reid 2002}
\tablenotetext{d}{6.0 $\pm$ 1.1 pc in Scholz et al.~2001; 5.2 $\pm$ 1.0 pc in Reid \& Cruz 2002}
\tablenotetext{e}{4.1 $\pm$ 0.6 pc in Delfosse et al.~2001}
\tablenotetext{f}{13.9 $\pm$ 3.5 pc in Lepine et al.~2002}
\tablenotetext{g}{7.2 pc and 29.5 pc in Reyle et al.~2002}
\tablenotetext{h}{7.6 pc in Henry et al.~2002}
\tablenotetext{i}{6.6 pc in Henry et al.~2002}
\tablenotetext{j}{11.9 $\pm$ 1.9 pc in Cruz \& Reid 2002}
\tablenotetext{k}{10.9 $\pm$ 2.2 pc in Reid \& Cruz 2002; 14.7 $\pm$ 0.4 pc in Cruz \& Reid 2002}
\tablenotetext{l}{8.5 $\pm$ 1.7 pc in Reid \& Cruz 2002}
\tablenotetext{m}{6.3 $\pm$ 1.3 pc in Reid \& Cruz 2002}
\tablenotetext{n}{10.9 $\pm$ 0.6 pc in Cruz \& Reid 2002}

\end{deluxetable}												  
}
\clearpage													  
														  
\end{document}